\documentclass[12pt,onecolumn]{emulateapj}

\newcommand{\M}{\mathrm}
\newcommand{\E}{\,\mathrm}

\newcommand{\magR}{{\M{mag}_\M{R}}}
\newcommand{\fluxR}{{\M{flux}_\M{R}}}

\def\hmpc{\ifmmode{h_{100}^{-1}\,\hbox{Mpc}}\else{$h^{-1}$\thinspace Mpc}\fi}
\newcommand{\iss}{{s}}

\newcommand{\Dol}{{D_\mathrm{ol}}}

\newcommand{\Dos}{{D_\mathrm{os}}}
\newcommand{\RE}{{R_\mathrm{E}}}

\newcommand{\tE}{{t_\mathrm{E}}}

\newcommand{\Int}{\int\limits}

\newcommand{\Intninf}{\int\limits_0^{\infty}}

\newcommand{\tfwhm}{{t_\mathrm{FWHM}}}

\newcommand{\vt}{{v_\mathrm{t}}}
\newcommand{\vtfs}{{v_\mathrm{t}^\ast}}

\newcommand{\pv}{{p_\vt}}

\newcommand{\pcmd}{{p_\M{cmd}}}

\newcommand{\sigl}{{\sigma_\M{l}}}
\newcommand{\sigs}{{\sigma_\M{s}}}
\newcommand{\sigls}{{\sigma_\M{ls}}}

\newcommand{\A}{{A_0}}
\newcommand{\uub}{{u_0}}

\newcommand{\F}{{F_0}}

\newcommand{\DF}{{\Delta F}}

\newcommand{\DFmax}{{\Delta F_\mathrm{max}}}
\newcommand{\Mlum}{{\mathcal{M}}}
\newcommand{\FVega}{{F_\M{Vega}}}

\newcommand{\ml}{{M}}
\newcommand{\Mlens}{{\ml_0}}

\newcommand{\Col}{{\mathcal{C}}}

\newcommand{\ext}{\mathcal{A}}

\newcommand{\xmeas}{x^\mathrm{meas}}
\newcommand{\ymeas}{y^\mathrm{meas}}
\newcommand{\DFmeas}{{\Delta_F^\mathrm{meas}}}
\newcommand{\Colmeas}{{\mathcal{C}^\mathrm{meas}}}
\newcommand{\tfmeas}{{t_\mathrm{FWHM}^\mathrm{meas}}}

\newcommand{\dudA}{\left|\frac{d\uub}{d\A}\right|}

\newcommand{\ufs}{{u_0^\ast}}
\newcommand{\bfs}{{b^\ast}}

\newcommand{\tfsFWHM}{{t^\ast_\M{FWHM}}}

\newcommand{\Rstar}{R_\ast}
\newcommand{\Msun}{{\ml_\odot}}

\newcommand{\Rsun}{{R_\odot}}
\newcommand{\RI}{(R-I)}

\shortauthors{Riffeser et al.}
\begin{document}
\title{The M31 microlensing event WeCAPP-GL1/Point-AGAPE-S3:\\
evidence for a MACHO component in the dark halo of M31?}

\shorttitle{WeCAPP-GL1: evidence for a MACHO component in M31?} 

\author{A.~Riffeser\altaffilmark{1}, S.~Seitz\altaffilmark{1,2} and R.~Bender\altaffilmark{1,2}}

\affil{University Observatory Munich, Scheinerstrasse 1, 81679 M\"unchen, Germany}
\affil{Max Planck Institute for Extraterrestrial Physics, Giessenbachstrasse, 85748 Garching, Germany}

   \email{arri@usm.lmu.de}

\begin{abstract}
  We re-analyze the M31 microlensing event WeCAPP-GL1/Point-AGAPE-S3
  taking into account that stars are not point-like but extended. We
  show that the finite size of stars can dramatically change the
  self-lensing event rate and (less dramatically) also the halo
  lensing event rate, if events are as bright as WeCAPP-GL1.  The
  brightness of the {brightest} events mostly depends on the source
  sizes and fluxes and on the distance distribution of sources and
  lenses and therefore can be used as a sensitive discriminator
  between halo-lensing and self-lensing events, provided the stellar
  population mix of source stars is known well enough. Using a
  realistic model for the 3D-light distribution, stellar population
  and extinction of M31, we show that an event like WeCAPP-GL1 is very
  unlikely to be caused by self-lensing.  In the entire WeCAPP-field
  ($17.2'\times 17.2'$ centered on the bulge) we expect only one
  self-lensing event every {49} years with the approximate parameters
  of WeCAPP-GL1 ({full-width-half-maximum (FWHM)} time-scale between 1
  and 3 days and a flux excess of {19.0~mag} or larger in $R$). On the
  other hand, if we assume only {20\%} of the dark halos of M31 and
  the Milky-Way consist of 1 solar mass MACHOs (Massive Astrophysical
  Compact Halo Objects) an event like WeCAPP-GL1 would occur every
  {10} years. Further more, if one uses position, FWHM time scale,
  flux excess and color of WeCAPP-GL1, self-lensing is even {13} times
  less likely than lensing by a MACHO, if MACHOs contribute {20\%} to
  the total halo mass and have masses in the range of 0.1 to 4 solar
  masses. We also demonstrate that (i) the brightness distribution of
  events in general is a good discriminator between self and halo
  lensing (ii) the time-scale distribution is a good discriminator if
  the MACHO mass is larger than 1 solar masses.  Future surveys of M31
  like PAndromeda (Pan-STARRS 1) should be able to provide many more
  such events within the next 4 years.
\end{abstract}


\section{Introduction}

Microlensing searches towards Local Group galaxies are interesting in
at least two respects. They can constrain the fraction of compact halo
dark matter (MACHOs, see
\cite{1986ApJ...304....1P,1991ApJ...366..412G}) and they allow to
study stellar populations and the 3D distribution of stars in the
Milky Way and the target galaxies.  Lensing events caused by stars
(self-lensing) also define a lower limit to the number of lensing
events that have to be identified in a survey and therefore provide a
consistency check for the lens model (depending on stellar population
content, stellar dynamics and density distribution of the stars) and
for the survey efficiency (see \cite{2001ApJS..136..439A} for MACHO,
and \cite{2007A&A...469..387T,2003A&A...404..145A} for EROS and
\cite{2005A&A...443..911C} for POINT-AGAPE and
\cite{2006A&A...446..855D} for MEGA). One can use the known
characteristics of lensing and self lensing events to design surveys
that will be dominated by self lensing. These self lensing-surveys can
measure the faint end mass function of stellar populations (see
\cite{2006ApJS..163..225R}).
\\
One can obtain the most likely MACHO-mass fraction and its confidence
limits from the analysis of all lensing events found in a survey
(using number, spatial distribution, time scale distribution etc.)
only after the selection criteria and the survey efficiency have been
taken into account. The survey efficiency depends on the event
characteristics (location, color, time scale and flux excess, finite
source effects etc.) and on the sampling and photometric quality of
observations. Including precise values for the survey efficiency can
completely change the interpretation of a survey.
\cite{2004astro.ph..8204P} concluded that there is no evidence for
MACHOs towards M31 in the INT data set, whereas the same collaboration
claimed with \cite{2005A&A...443..911C} that there is a fairly strong
evidence based on new efficiency estimates of the same survey.
\\
Instead of comparing the expected and observed events one can analyze
the observables of individual lensing events; these are flux excess at
maximum magnification, full width half maximum time, color, location
and presence/absence of finite source signatures in the light curve
(see \cite{2006ApJS..163..225R} for more details). One can ask for the
relative probability of halo lensing and self lensing and derive
probability distributions for the lens masses causing that event.
This has been done in a simplified way for WeCAPP-GL1
(\cite{2003ApJ...599L..17R}, or \cite{2003A&A...405...15P} for the
POINT-AGAPE-S3 identification of the same event) and also for
POINT-AGAPE N1 \citep{2001ApJ...553L.137A} and N2
\citep{2004ApJ...601..845A} in the past.  However, the sources for all
three events have been treated as point sources. This can mislead the
interpretation, in particular if the events are very bright.  {This is
  also the case for light curves not showing any finite source
  signatures because the number of possible lens-source configurations
  strongly changes for bright events and the interpretation of the
  lens mass is modified.}
\\
It is the subject of our paper to (re)-analyze WeCAPP-GL1 and derive
the relative probabilities of self lensing and halo lensing accounting
for the extended sizes of stars.  We summarize our M31 model in
Appendix \S\ref{model} and present the equations needed for extended
source stars in Appendix \ref{app.eventrates} and
\ref{sec.analy_massprob}.
{We demonstrate the pronounced differences resulting from this more
accurate description of M31 in \S\ref{sec.improvedmodel}.}
In \S\ref{sec.pointsource} we present the
lens mass distribution without accounting for finite source effects
and in \S\ref{sec.qualitative} we show some qualitative results using
the radius for the source stars. We evaluate in
\S\ref{sec.quantitative} an accurate estimate for the mass
distribution of the event using the stellar source size.  We
investigate in \S\ref{sec.extinction} how sensitive the results are
with respect to the {mean} M31 extinction and the 
  line-of-sight (LOS) extinction towards WeCAPP-GL1 and in
\S\ref{sec.youngdisk} with respect to the stellar population
properties of the disk stars acting as sources.  In
\S\ref{sec.statistics} we give an outlook to the statistical
interpretation for microlensing events in M31. Finally we draw our
conclusions in \S\ref{conclusions}.

\section{An improved WeCAPP-GL1 analysis in the point source
    approximation}
\label{sec.improvedmodel}

\begin{figure}
  \centering
  \includegraphics[width=0.65\textwidth]{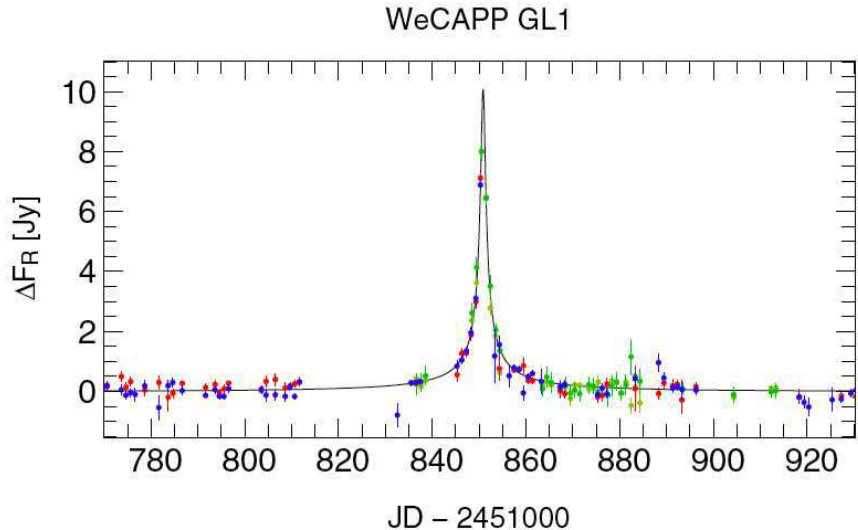}
  \caption{{Light curve of WeCAPP GL1 (R band: blue, I band transformed to R band: red); POINT-AGAPE-S3 (r' band: dark green, i' band: light green)}}
  \label{fig.GL1}
\end{figure}

{For the current analysis we improve our light curve fit for
  WeCAPP-GL1 (Fig.~\ref{fig.GL1}) with respect to
  \cite{2003ApJ...599L..17R}, Table 2 by using a Paczy\'nski
  microlensing light curve fit (5+2+2+2 free parameters in 4 filters; $t_0$, $u_0$, $t_\M{E}$, 
  $F_{0,\M{R}}$, $C_\M{R}$, $F_{0,\M{I}}$, $C_\M{I}$, $F_{0,\M{r'}}$, $C_\M{r'}$, $F_{0,\M{i'}}$, $C_\M{i'}$)
  for the observables and a Gould microlensing function to determine
  the errors of the observables $\tfwhm$ and $\DF$.  We also improved
  our photometric calibration accounting for the color terms in the
  filter calibration; our instrumental magnitudes $\tilde{R}$ and
  $\tilde{I}$ [phot/sec at AM=0] transform to the Johnson system with
  $R=\tilde{R} + 23.58 - 0.01 (\tilde{R}-\tilde{I})$ and $I=\tilde{I}
  + 22.89 + 0.22 (\tilde{R}-\tilde{I})$.  The resulting parameters are
  listed in Table~1.  Note that the fit implies that the
  source star of GL1 is a bright Red Giant Branch (RGB) star.}

\begin{table*}[h]
  \scriptsize
  \setlength{\tabcolsep}{1mm}
  \centering
  \begin{tabular}{|c|c|c|c|c|c|c|c|c|c|c|c|}
  \hline
     WeCAPP GL1 & $\alpha$    & $\delta$   & $t_0$        & $\tfwhm$ & $\Delta F_\M{R}$ & $\Delta F_\M{I}$  & $\RI$ & $A_0$ & $\Mlum_R$ & $\chi_\M{dof}$ \\
               &  (2000)     &   (2000)   & (JD-2450000) &      [d] & [$10^{-5}\E{Jy}$] &  [$10^{-5}\E{Jy}$] & [mag] &         & [mag] &    \\              
  \hline  									
  observables  & $00^h 42^m 30.28^s$ & $41^\circ 13' 01.1''$ &     2451850.86 & 1.83  & 10.07 & 17.12 & 0.83 &  108 & -0.64                & 1.33 \\
  error        & $\pm 0.2''$             & $\pm 0.2''$              &           $\pm 0.02$ & $\pm 0.10$  &  $\pm 0.44$ &  $\pm 0.78$ & $\pm 0.03$ &   $\pm 57$ & $-0.46/+0.82$ &     \\
  \hline
  \end{tabular}								 
  \caption{Summary of WeCAPP-GL1 (or
    AGAPE-S3) observables ($\alpha$ and $\delta$) and parameters derived
from a Paczy\'nski { microlensing light curve} fit: date of light curve maximum,
    full-width-half-maximum time $\tfwhm$ of the event, flux excess at
    light curve maximum in R and I-band, color of event, quality of
    the fit, and an estimate for the degenerate amplification and absolute brightness of the source without errors.}  
\end{table*}

Despite the well-sampled light curve of the event and the nicely
fitting Paczy\'nski light curve, finite source signatures above the
highest data point of $8\times 10^{-5}\E{Jy}$ cannot be ruled out. 
{Moreover, as we will demonstrate later on,} finite source
signatures for events brighter than $19\E{mag}$ are indeed more likely
than no finite source signatures. {In a future work we are
planning to use a modified light curve (see
\cite{2006ApJS..163..225R}, Eq.~12) that allows to include the
finiteness of lenses and sources and their limb darkening.}

The mass probability function for WeCAPP-GL1 was already derived in
{\cite{2003ApJ...599L..17R}} using a simplified analysis of the event
and a simple description of M31.
\\
We now use the improved analysis method derived in \cite{2006ApJS..163..225R}
and { a more detailed M31 model}.  We use this M31 model for all
calculations further on {unless stated otherwise}.
The relevant
differences for the analysis of WeCAPP-GL1 are as follows:

\begin{enumerate}
\item {\bf Color-magnitude relation of source stars}\\
  In {\cite{2003ApJ...599L..17R}} we approximated the
  color-magnitude-relation $\pcmd_\iss(\Mlum,\Col)$ of bulge and disk
  stars in M31 with observations of M32 to derive a brightness
  estimate for {a} post main sequence star with a color of the
  WeCAPP-GL1 event. In this way, the brightness of a bulge star and a
  disk star with a color of WeCAPP-GL1 was estimated the same,
  $M_R=-1.5\E{mag}$.  

  We now model the disk and bulge population separately, using the
  stellar population isochrones of \cite{2002A&A...391..195G} with a
  metalicity of $Z=Z_\odot$ ({\tt isoc\_z019.dat}).  We describe the
  bulge as a 12.6 Gyr single stellar population (SSP).  

  {The disk is modeled as a composite of 6 SSPs with ages of 4,
    20, 100, 500 Myrs and 2.5 and 12.6 Gyrs. Their relative weights
    are calculated by integrating an exponentially declining star
    formation rate with $\tau=8\E{Gyrs}$ over adjacent intervals (see
    Table~2).  This simple model is close to a population where stars
    formed continuously over the past 12 Gyrs. 
    In Figs.~\ref{fig.CMD} and \ref{fig.mass_function} 
    we compare the luminosity and mass function of this composite
    stellar population with a population composed of 71 bursts with
    the same star formation decline rate of 8 Gyrs. The differences are
    marginal, i.e., our `simple' model describes the brightness and
    mass distribution of stars with an e-folding timescale of 8 Gyrs
    fairly well, and offers the advantage of much faster numerical
    integrations in the color-magnitude plane.}

\begin{table*}[h]
  \centering
  \begin{tabular}{l|l|l}
    age [Myr] &   interval [Myr] &  weight \\  
    \cline{1-3}
          4.0 &    0.0 -    12.0 & 0.00039 \\
         20.0 &   12.0 -    60.0 & 0.00158 \\
        100.0 &   60.0 -   300.6 & 0.00804 \\
        501.2 &  300.6 -  1506.5 & 0.04417 \\
       2511.9 & 1506.5 -  7550.6 & 0.35630 \\
      12589.3 & 7550.6 - 12589.3 & 0.58952 \\
  \end{tabular}		
  \caption{}						 
\end{table*} 
  The {star formation history} of M31 is not known exactly\footnote{
    \cite{2003ApJ...592L..17B}'s data and analysis suggest that (at
    least at a distance of $10\E{kpc}$ where their data have been
    taken) about 30$\%$ percent of stars (in mass) could be as young
    as $6-8\E{Gyr}$. {Also, the metalicity is large} [see
    \cite{1986ApJ...305..591M}, {[M/H] =-0.6}
    \cite{1994AJ....108.2114D}, {[Fe/H]=-0.6}
    \cite{1996ASPC...92..544R}  {[Fe/H]=-0.4}
    ] and falls off only at more than $30\E{kpc}$
    \cite{2006ApJ...648..389K}.}.  Our choice ensures the presence of
  mostly old stars with a small fraction of young, blue and bright
  stars.  

  {This implies that the stars in our model also populate the
  extremes of the M31-CMD (see Fig.~\ref{fig.CMD}).}

We thus can allow for the
  effect that a few bright stars might be more efficient sources for
  detectable microlensing events than many fainter, older stars.
  \\
  We now do not have to estimate the {typical} stellar brightness at a
  given color anymore, but can construct a brightness-color
  probability distribution $\pcmd(\Mlum,\Col)$ from the isochrones and the assumed stellar
  mass function $\xi(\ml)$. The information provided in the Girardi isochrones
  also allows a straightforward extraction of the stellar radii $\Rstar(\Mlum,\Col)$, which
  is necessary for an accurate treatment of finite-source size
  effects.
\\
  Note that the color information of the event helps to select a
  smaller set of sources and therefore the mass probability results
  from a smaller range of source brightness. Because of this
  additional information the mass probability distributions become
  narrower and more precise.

\begin{figure}
  \centering
  \includegraphics[width=0.8\textwidth]{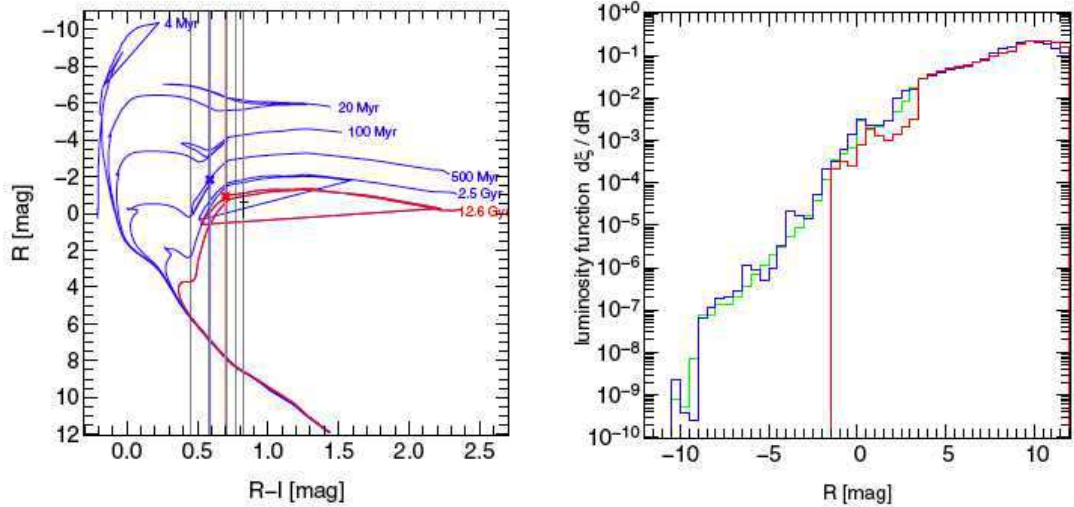}
  \caption{{{\it Left panel:} Color $\RI$ -- $R$-band magnitude
    relations for stellar populations of the bulge ({\it red}) and disk
    components ({\it blue}) with solar metalicities (using isochrones of
    \cite{2002A&A...391..195G} in the Johnson filter system). 
    {\it Right panel:} Luminosity function $\xi$for the bulge 
    ({\it red}) and disk ({\it blue}) combining isochrones, weights and mass functions of our model. The {\it green} curve shows a nearly continuous disk population with 71 components.}}
  \label{fig.CMD}
\end{figure}

\item {\bf Analysis method}\\
  In \cite{2003ApJ...599L..17R} we had estimated the approximate
    source brightness $\Mlum$ from the color of the event $\Colmeas$, and
    derived a magnification $\A$ to convert the observed
    full width half maximum time of the event $\tfmeas$ to its
    Einstein-time scale   
  $\tE\approx\tfmeas/\sqrt{12} (\A-1) \approx \tfmeas/\sqrt{12} \,\DFmeas/\{\FVega
    10^{-0.4[\Mlum(\pcmd,\Colmeas)+24.43]}\}$.
\\
Using descriptions for bulge/disk/halo densities $\rho$ and velocities $\pv$, for the
mass functions of stars $\xi(\ml)$ and a single mass MACHO halo we
obtained the relative probabilities $p(\ml)$ for self lensing and halo
lensing (as a function of the MACHO mass) with
\begin{equation}
  p(\ml,\tE)\sim \,\xi(\ml) \Intninf \rho_s(\Dos) \Int_0^\Dos
  \rho_l(\Dol) \;\;\pv{\left(\frac{\RE}{\tE}\right)} 
  \;\left(\frac{\RE}{\tE}\right)^3 \,d\Dol\,
  \,d\Dos \quad.
\end{equation}

This means that the lens mass probability distribution $p(M,\tE)$ was
obtained using the (unobservable) maximum magnification $\A$ and
Einstein-time $\tE$ instead of the observables flux-excess $\DFmeas$
and full-width half maximum time scale $\tfmeas$.

We have shown
  in \cite{2006ApJS..163..225R} (Eqs. 80 and 84) that the
  Einstein-time--impact-parameter degeneracy leads to coupled errors
  for these two quantities, and thus should be avoided. In the
  following we will always use the {real} observables flux excess $\DF$
  and time-scale $\tfwhm$ instead of maximum magnification and Einstein time.
{Using the notation of \cite{2006ApJS..163..225R}, the event rate per
area, per event time scale $\tfwhm$, per flux excess $\DF$, per source
color $\Col$, per absolute magnitude of the source star $\Mlum$, and
per lens mass $\ml$ is
\begin{equation}
\frac{d^7 \Gamma(x, y, \tfwhm, \DF,  \Col ,\ml, \Mlum)}{\,dx \,dy
   \,d\tfwhm \,d\DF \,d\Col  \,d\ml \,d\Mlum } = 
   \frac{2 \,\pcmd \,\xi}{\tfwhm^3} 
   \Intninf n 
    \left( \frac{\Psi}{\F}   \Int_0^{D_\M{ol}^\ast}
	 \rho  \,R_\M{E}^{3} 
    \,\pv(\vt) \,d\Dol \,+ 
  \,\Omega^\ast
    \, \rho^\ast \,R_\M{E}^{\ast 3} 
		 \Int_0^{u_0^\ast} 
		 \,\pv(v_\M{t}^\ast) 
		 \,{\Upsilon^\ast}^2
		  \,d\uub
\right) d\Dos
\label{eq.eventrate}
\end{equation}

All definitions are explained in Appendix~\ref{app.eventrates}.}

\item {\bf Mass function of stars} \\
  \cite{2003ApJ...599L..17R} assumed that both disk and bulge
  stars are confined to a mass interval from $0.08\Msun$ to $0.95\Msun$
  (bulge) and to $10\Msun$ (disk) with a \cite{2000ApJ...530..418Z}
    IMF ($\xi \sim M^{-1.33}$) for the bulge and a
    \cite{1997ApJ...482..913G} IMF for the disk, respectively.  
    We now use those mass functions which are consistent with the stellar population
    models, i.e., a combination of a \cite{2007A&A...467..117B} \&
    \cite{2002Sci...295...82K} IMF for the bulge and a
    \cite{1997ApJ...482..913G} \& \cite{2002Sci...295...82K} IMF for
    the disk. Mass loss is provided by stellar
    population models \citep{2002A&A...391..195G} and is taken into account in the mass function (MF).

    We also extended the lower mass limits to $0.001\Msun$.  

    Because of the small gradient of the MF for very small masses {and
      their low cross section for microlensing} all results are
    independent from the lower mass limit.

    We also included stellar remnants originating at the high mass end
    of the IMF into the MF (see \cite{1993ApJ...416L..49R}).

\begin{figure}[h]
  \centering
  \includegraphics[width=0.6\textwidth]{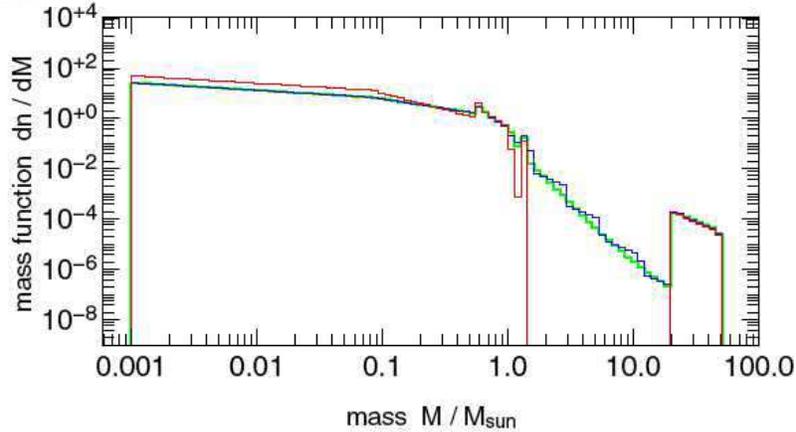}
  \caption{Present day mass function (MF) for M31 bulge ({\it red}) and
      disk ({\it blue}) population (mass loss according to \cite{2002A&A...391..195G}). 
      Stars more massive than $1\Msun$ are
      either young disk stars ({\it blue}) or disk/bulge stellar remnants
      (white dwarfs $\le 1.13\Msun$, neutron stars $\approx 1.4\Msun$,
      black holes $\ge 20\Msun$). {The {\it green} curve shows a nearly continuous disk population with 71 components.}}
  \label{fig.mass_function} 
\end{figure}

\item {\bf Extinction} \\ 
  In \cite{2003ApJ...599L..17R} we did not account
  for the extinction along the line of sight to the source of the
  lensing event.  Extinction alters the {true} color and brightness
  and thus the necessary magnification of a lensed source, { which}
  changes the shape of the mass probability function extracted from
  individual events.  

  {The overall extinction decreases the observed flux, leading to
    the need} to increase the number of stars (and thus the mass) to
    obtain the same luminosity.

  Extinction in this way also influences the absolute and
  relative amplitudes of lensing and self lensing rates.  The model of
  \cite{2006ApJS..163..225R} assumes an {on average} M31-extinction
  for all disk stars in the WeCAPP field of $0.51$ mag and for all bulge
  stars of $0.19$ mag in the R-band, independent of the angular position
  of the event, and independent of the LOS distance to the source.
  The MW-extinction is added uniformly with $0.17$ mag (R-band). 

  With these assumptions on extinction one gets the following
  extincted colors for the bulge and disk stellar populations
  described above: for the bulge 
  $(U-B)_\M{meas} = 0.78$,
  $(B-V)_\M{meas} = 1.19$,
  $(B-R)_\M{meas} = 1.91$,
  $(V-R)_\M{meas} = 0.72$,
  $(R-I)_\M{meas} = 0.75$, 
  and for the disk
   $(U-B)_\M{meas} = -0.06$,
   $(B-V)_\M{meas} = 0.73$,
   $(B-R)_\M{meas} = 1.33$,
   $(V-R)_\M{meas} = 0.60$,
   $(R-I)_\M{meas} = 0.69$.
  Especially the values for the disk {agree well with} 
  \cite{1987A&AS...69..311W}.
  We will use these extinction values {if not specified otherwise}. 
    Additionally we use slightly refined extinction descriptions in
  \S~\ref{sec.extinction} later on.
\item {\bf Mass-to-light ratio $(M/L)$ and total mass of bulge, disk and halo} \\
  { Combining the luminosity function and the mass function results
    in the $M/L$ ratios of $(M/L)_R= 3.3$ (bulge) and
    $(M/L)_R= 1.2$ (disk) not including extinction.

    Using these $(M/L)$ with the M31-light-model and the
    extinction assumptions results in a very similar massive bulge
    ($4.4\times10^{10}\Msun$) and a more massive disk
    ($4.2\times10^{10}\Msun$). Since we changed in particular the disk
    mass we now use a different halo model with a core radius of 5 kpc,
    a total mass of $1.23\times10^{12}\Msun$, and a cut-off radius of
    100 kpc.  Note that the halo model has a large uncertainty since a
    large set of different combinations of core-radius,
    central-density, cut-off radius are able to fit the measured
    rotation curve of M31.}

\end{enumerate}

\section{Point source approximation}
\label{sec.pointsource}

In this {section} we present the lens mass distribution without
accounting for finite source effects.  We now demonstrate the impact
of the change of assumptions described above on the interpretation of
the event relative to \cite{2003ApJ...599L..17R}.  In this chapter the
sources are still treated as a point source.

\begin{figure}[h]
  \centering
  \includegraphics[width=0.8\textwidth]{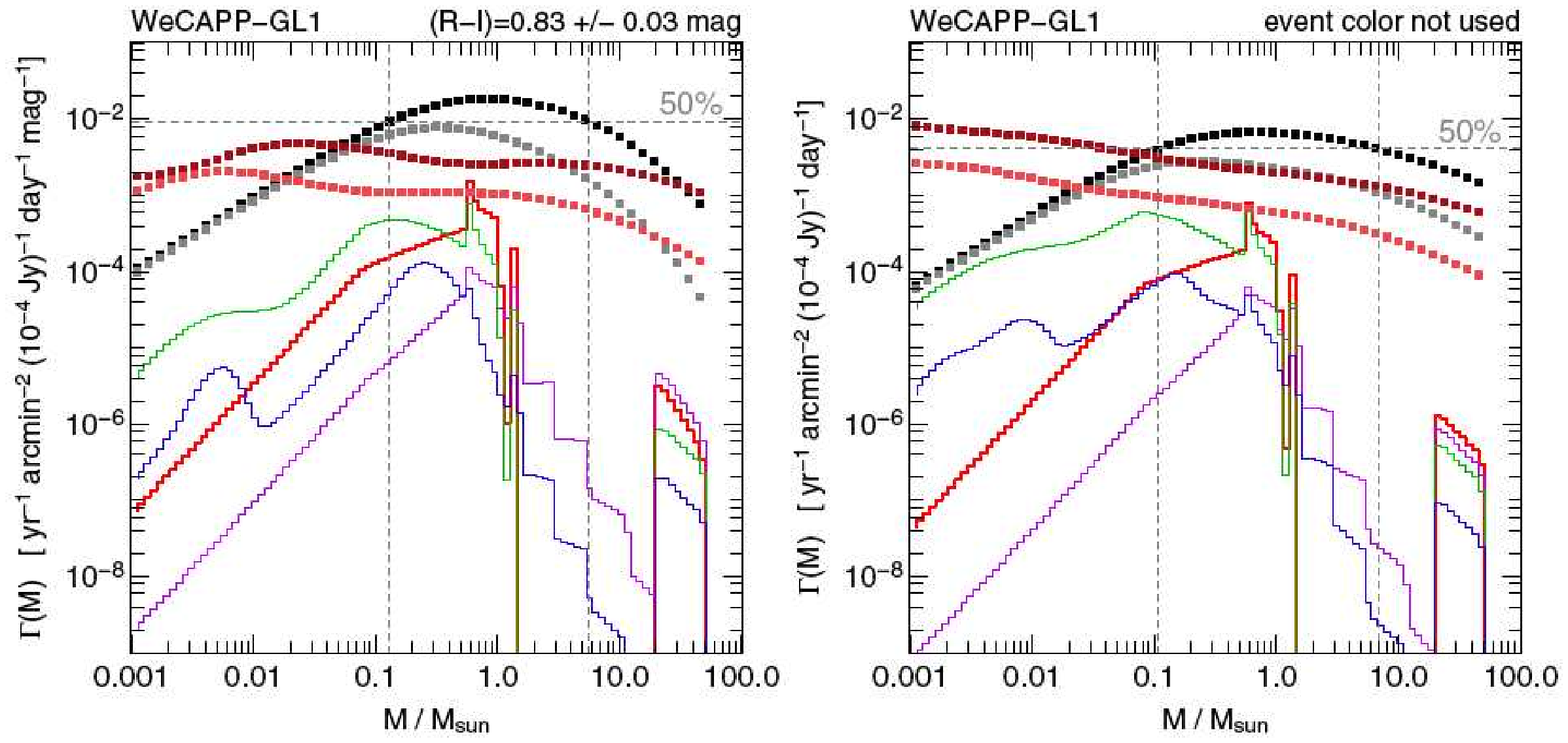}
	\caption{Lens mass distribution [event rate $\Gamma(\ml)$ per area,
    per event brightness $\DF$, per event time scale $\tfwhm$, per
    color $\RI$] for WeCAPP-GL1 in the {\bf point source
      approximation} using LOS position ($\xmeas$,$\ymeas$),
    time-scale ($\tfmeas$), and flux-excess ($\DFmeas$) and their
    errors (Eq.~\ref{eq.eventrate_obs_6dim}). {In the {\it left
        panel} the measured event color ($\Colmeas$) was
      transformed to the un-reddened color $(R-I)_0=0.59$ for disk sources, and
      $(R-I)_0=0.70$ for bulge sources, both with a Gaussian error of
      $0.03$. The {\it right panel} was calculated not using the
      color information of the event.}
   The {\it red}, {\it green}, {\it blue} and {\it purple} curves
    show the event rate for bulge-bulge, bulge-disk (bulge lens and
    disk source), disk-bulge and disk-disk lensing.  
    The mass range is confined to the MF interval, and
    therefore there is an upper limit of about $50\Msun$ resulting
    from black hole remnants.  
{The amplitudes of the lens mass distributions are scaled such,
    that their maxima represent the expected rate due to the different
    lensing and self lensing scenarios and the ratios of their maxima
    yield the relative probabilities for the different scenarios.
    The information about halo lensing (M31-halo-bulge and
    M31-halo-disk data points are in {\it black} and {\it grey}, and
    MW-halo-bulge and MW-halo-disk in {\it brown} and {\it
      light-brown}) is displayed as points and not by a curve: we
    assume a {mono-mass-spectrum} for the halos, and each point
    represents a halo which is made of a given MACHO-mass to 100\%.
    For each MACHO mass one obtains the event rate from the amplitude
    at that mass.  } }
  \label{improved_pointsource_p_of_M} 
\end{figure}

{The left panel of} Figure~\ref{improved_pointsource_p_of_M} shows
the lens mass distribution (event rate $\Gamma(\ml)$ per area, per
event brightness $\DF$, per event time scale $\tfwhm$) for WeCAPP-GL1
in the point source approximation using the LOS position
($\xmeas$,$\ymeas$), the time-scale ($\tfmeas$), the flux-excess ($\DFmeas$) and the
color ($\Colmeas$) of the event and their errors.  The assumptions in
evaluating the event rates have been changed relative to
\cite{2003ApJ...599L..17R} as detailed in (1) to (5).
{The event rate for bulge-bulge lensing is 2 times as likely as
  bulge-disk lensing, about 13 times as likely as disk-bulge lensing
  and about 12 times as disk-disk lensing.
  For lensing of a bulge star by a $0.74\Msun$ MACHO in M31 we
  obtained an event rate which is 12 times as likely as bulge-bulge
  self lensing. Using the event color the most likely halo lens masses
  for WeCAPP-GL1 are between $0.13\Msun$ and $5.6\Msun$.}
The main difference of this new analysis (still in the point source
approximation!) relative to the result of \cite{2003ApJ...599L..17R}
is that the M31-halo-disk lensing scenario becomes now up to a factor
{1/4} similar to the halo-bulge lensing (compared to
\cite{2003ApJ...599L..17R}, where the M31-halo-bulge lensing was about
10 times as likely as halo-disk lensing).
For the point source approximation the total halo-lensing contribution
(summing over M31 and MW halo-bulge and halo-disk lensing) is {13} times
more likely than all self lensing contributions.

{In the right panel of
  Figure~\ref{improved_pointsource_p_of_M} we show the same
  calculations without using the color information of the event.}
At low masses the curves for $\Gamma(\ml)$ are only changed for disk
sources, whereas bulge sources are not affected.  The high mass end
cut-off however is suppressed for all lens-source configurations due
to the use of the measured event color. This is because the color of
the WeCAPP-GL1 makes the source most efficient for becoming a very
luminous event, and sources with colors different from WeCAPP-GL1 have
smaller fluxes requiring larger Einstein radii and thus larger lens
masses to achieve an event of the same brightness. Therefore, dropping
the color information for WeCAPP-GL1 mainly affects the high mass
probabilities.
%
\begin{figure}[h] 
  \centering
  \includegraphics[width=0.98\textwidth]{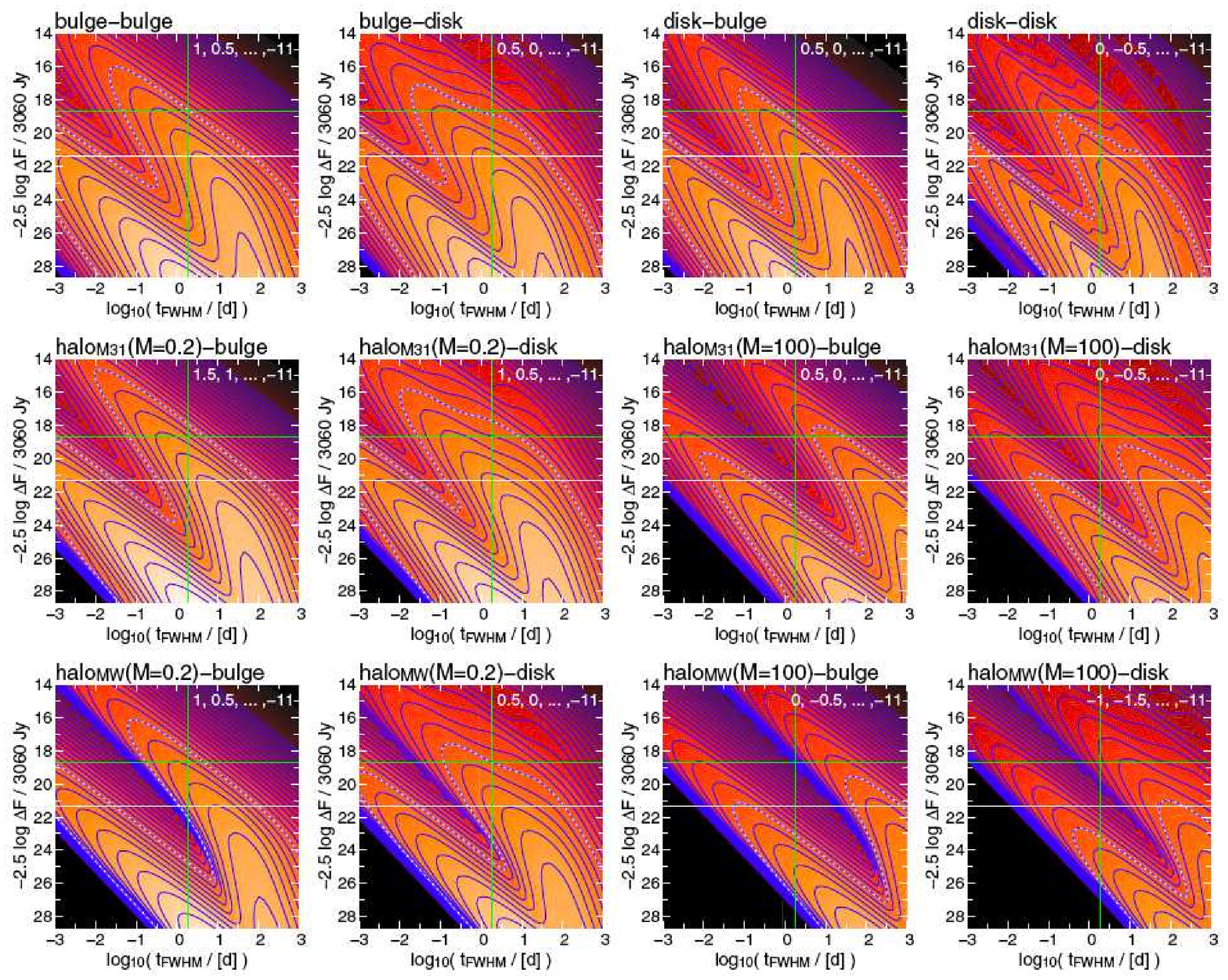}
  \caption{Contour plot of the time-scale and
    flux-excess distribution ($\DF$-$\tfwhm$-plane) of events at the
    location and with the intrinsic color of WeCAPP-GL1,
    [$(R-I)_0=0.59$ for disk sources, $(R-I)_0=0.70$ for bulge sources,
     without using its error], obtained from
    Eq.~\ref{eq.eventrate_obs_5dim_log} in the {\bf point source
      approximation}. The first row shows self lensing, the second
      row M31 halo-lensing, {the third
      row MW halo-lensing}. The two further observables of WeCAPP-GL1
    (flux excess and event time scale obtained from a point source
    lensing light curve fit) are marked in {\it green}.  { The
      estimate for the WeCAPP 6$\sigma$ detection limit (see
      \cite{2006ApJS..163..225R}, Table 1) at the position of GL1 is
      shown as {\it white line}.}  The contours differ { by factors of}
    $10^{0.5}$, and the dashed contour is that of $10^{-3}$ events
    per year, per square arc minute at { the} location of WeCAPP-GL1,
    per color (in magnitudes), per logarithmic timescale (time
    scale in days) and per flux-excess (converted to magnitudes) 
    (see footnotes~\ref{fn.mag_flux} and \ref{fn.conversion}). }
  \label{fig.events_tf_DF_all} 
\end{figure}

In Fig.~\ref{fig.events_tf_DF_all} we show the distribution of events in
the $\DF$-$\tfwhm$-plane in the point source approximation as obtained
from Eq.~\ref{eq.eventrate_obs_5dim_log} with location and intrinsic
color of WeCAPP-GL1, [$(R-I)_0=0.59$ for disk sources, $(R-I)_0=0.70$ for
bulge sources,  without using its error]. 
Since in the point source approximation arbitrary large magnifications
are possible, there is no limit for the brightness or shortness of the
events. The event rate for events like WeCAPP-GL1 is equal for
bulge-bulge and bulge-disk lensing. MACHO-lensing events caused by a
$0.2\,\Msun$ MACHO halo are roughly a factor of 10 more likely
than self lensing for a full MACHO-halo.

Here the color information affected the shape of the event-rate
contours in the $\tfwhm$-$\DF$-plane.  
The {`tilted M'}-shape comes from the bimodal luminosity distribution of
stars with color of WeCAPP-GL1:
The luminosity function maps the magnification-$\tfwhm$-distribution
into the $\DF$ space, see Eq.~62 in \cite{2006ApJS..163..225R}.
The color of the event constrains the luminosity of the {post main
  sequence (PMS)} sources
fairly well, and one thus gets the {`tilted M'}-structure of the event
rate in the $\DF$-$\tfwhm$-plane. 
The left part is due to main sequence source stars, the right part due
to {PMS} stars. Main sequence stars need larger
magnification for the same event brightness, and since large
magnification needs small impact parameters $b$, this goes in parallel
with short FWHM-timescales, $\tfwhm \propto b/\vt$, of the events.
If the PMS sources have a broader brightness distribution, the right
part of the {`M'} is washed out by moving it up and down vertically.
Integrating over all colors makes the {'M-shape'} almost disappear,
since the {width} of the {`M'} depends on the color (see
\cite{2006ApJS..163..225R} for more details). 
Since the intrinsic color of WeCAPP-GL1 for bulge sources is between
{$0.6$ and $0.8$} (depending on the degree of reddening, see chapter
\ref{sec.extinction}), and the magnitude of the PMS-stars is fairly
constant in this color range (see Fig.~\ref{fig.CMD}), the location of
the right part of the {'M'} is independent from color within the
possible range of intrinsic colors of WeCAPP-GL1.  The location of the
event relative to the {`tilted M'} provides an indication weather the
time scale of the event fits to its brightness.

{Figure~\ref{fig.events_tf_DF_all} can also be used to visualize
  how plausible the observables $\DF$-$\tfwhm$ of an event candidate
  are. E.g., if MACHOs would have 100 solar masses, then for a given
  brightness of GL1, its timescale would be too low -- or, on the
  other hand assuming GL1 is a convincing event, for MACHOs with
  $100\Msun$ one would expect many more events with the same
  brightness but timescales of 10 days (which are not found). }

{To summarize in the point source approximation an event as bright
  as WeCAPP-GL1 is quite unlikely to be a self lensing event. The most
  likely self lensing scenario is bulge-bulge self lensing.  The color
  information does only marginally influence the mass distributions.
  Noticeable is that for self-lensing low lens masses of $0.1\Msun$
  achieve relative high probabilities.  Self-lensing is half as likely
  as halo-lensing, if MACHOs contribute {20\%} to the total halo mass
  and have a mass of about $1\Msun$.}

\section{Effects of finite source sizes: a qualitative understanding
  of WeCAPP-GL1}
\label{sec.qualitative}

{In this chapter we show for the fittings parameters of the source star (color $(R-I)_\M{meas}=0.83$ and brightness $\Mlum_\M{meas}=-0.64$)
that the lensing interpretation is strongly influenced by the finite-source treatment.
{Here we discuss the different
competing effects qualitatively, while} the full quantitative formalism
is described in \S\ref{sec.quantitative}.

If a source is treated as point-like, it can be magnified at light
curve maximum by a foreground point mass with any magnification value
between one and infinity: for each Einstein-radius of a potential
lens, there exists a source-lens-trajectory (an impact parameter) such
that the desired magnification of the event can be achieved. However,
if the source is extended, then the magnification at light curve
maximum is no longer unlimited: 

\begin{itemize}
\item if the impact parameter is larger
than the projected source radius [$b>\bfs \approx \Rstar \Dol / (2
\Dos)$], nearly no finite source signatures appear in the event light curve,
and all relations for the point source approximation hold.

{\item if the impact parameter is closer ($b<\bfs$), the maximum
  magnification does only weakly depend on the impact parameter
  anymore. In our approximation (see \cite{2006ApJS..163..225R},
  Fig.~1) the maximum magnification depends solely on the
  Einstein-radius and on the size of the source star, projected onto
  the lens plane, $\Rstar\Dol/\Dos$ (see below,
  Eq.~(\ref{eq.DeltaFmax_FS})).  Therefore, for the highest
  magnifications there is no longer a trade-off between source-lens
  impact parameter $b$ and Einstein-radius $\RE$, but between
  projected source size $\Rstar\Dol/\Dos$ and Einstein-radius $\RE$.
  This means that the larger the projected source size is, the larger
  the Einstein-radius has to be to achieve a high magnification.}
\end{itemize}

 For self lensing, the Einstein-radii are
limited by the maximum stellar masses (e.g. around $1\Msun$ for the
M31 bulge population) or {the maximum} remnant masses, the inefficient lensing
geometries (fairly similar source $\Dos$ and lens distances $\Dol$),
and large projected source radii.  Therefore, high magnification
events are hard to achieve.  These effects are much less severe for
halo lensing events.  One therefore expects for bright lensing events
that halo lensing becomes more likely relative to self lensing if the
finite stellar sizes are taken into account. This could make the self
lensing hypothesis for an event as bright as WeCAPP-GL1 very unlikely,
although not completely un-feasible.

How {easily} an event can occur depends on the number of source-lens
pairs that can {produce} the event. The LOS density distributions
of bulge and disk are confined to a few kilo parsecs (see
Fig.~\ref{fig.CMD_GL1}, left) which implies that the
source-lens distances along the LOS must not be larger than
a few kilo parsec for self lensing; otherwise either the source or the
lens density and thus their product is small.

The relation between maximum magnification $\A=(\DF+\F)/\F \approx
\DF/\F$, Einstein-radius $\RE$, and projected source radius $\Rstar
\Dol / \Dos$ is (\cite{2006ApJS..163..225R}, Eq.~(18))

\begin{equation}
  \DFmax  =  F_0 \left(\sqrt{1+\frac{16 G \ml\,\Dos (\Dos-\Dol)}{c^2\,\Rstar^2 \,\Dol}}
    -1\right) 
  \approx 
  \frac{4\sqrt{G\Msun}}{\Rsun c}
  \,\sqrt{\frac{\Dos(\Dos-\Dol)}{\Dol}}
  \; \frac{F_0\;\sqrt{\ml/\Msun}}{\Rstar/\Rsun}
  \approx
2\F\frac{\RE}
{\Rstar\frac{\Dol}{\Dos}}\,\,, 
\label{eq.DeltaFmax_FS}
\end{equation}
where the un-lensed source flux $\F= \fluxR(\Mlum+\ext) (10\E{pc}/\Dos)^2$ (see footnote \ref{fn.mag_flux}) depends on the
absolute source luminosity $\Mlum$, the source distance $\Dos$ and the extinction  $\ext$
towards the source. {Note that our approximation Eq.~\ref{eq.DeltaFmax_FS} represents 
also the largest flux excess 
achievable by a light curve not showing finite source signatures.
}

Eq.~\ref{eq.DeltaFmax_FS} can  be inverted to
\begin{equation}
  \begin{array}{rcl}
		\Dol_\M{,max}
    \approx
    \Dos \left(1+\left(\frac{\Rsun c}{4 \sqrt{G \Msun}} \DFmax 
        \frac{\Rstar/\Rsun}{\F} \frac{1}{\sqrt{\ml/\Msun}} \right)^2 \frac{1}{\Dos} 
		\right)^{-1} 
    \quad.
  \end{array}
	\label{eq.Dol_GL1}
\end{equation}
Eq.~\ref{eq.Dol_GL1} provides the maximum distance a lens can
have (to us) to allow a lensing event with a flux-excess $\DFmax$ at
maximum magnification once the finite source sizes are taken into
account. For a given source distance and a given source and lens
population, the upper limit of that distance is set by the largest
{$\F/\Rstar$}-ratio for source stars and the largest lens mass
possible.
For a given (measured) flux excess of an event, Eq.~(\ref{eq.Dol_GL1})
depends on the extinction along the line of sight: the smaller the
extinction, the larger is $\DFmax$; also, the
radii $\Rstar$ have to be taken into account using the de-reddened
color of the event.  If the color information is available, one has to
use the stellar radii of the de-reddened source to obtain the
largest {$\F/\Rstar$} - ratio for source stars.

\begin{figure}
  \centering
  \includegraphics[width=0.9\textwidth]{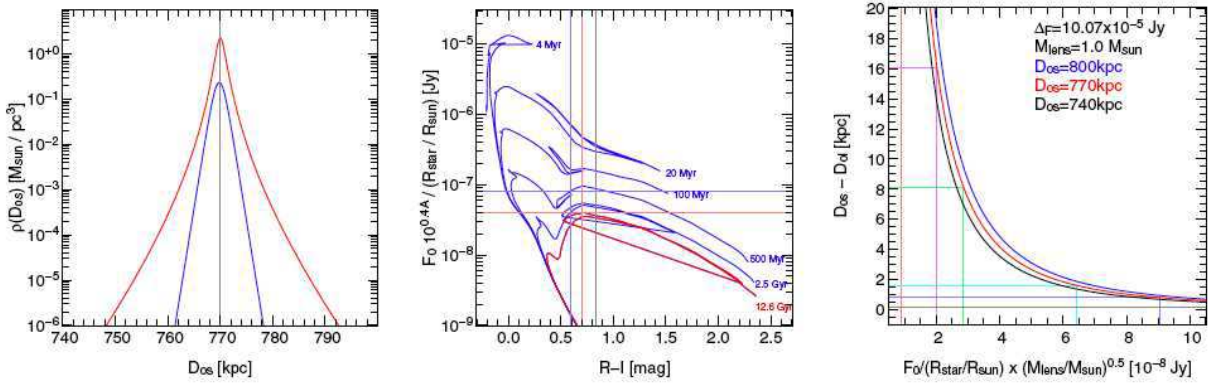}
  \caption{{\it Left panel} Density of bulge ({\it red}) and disk ({\it blue}) 
    stars along the
    LOS to WeCAPP-GL1, using the M31-model described in
    \cite{2006ApJS..163..225R}.
    In the {\it middle panel} we have transferred the stellar
    brightness to a flux (in Jansky's) at 770kpc (distance of M31), and
    obtained the ratio of this flux to the stellar size (in solar
    units). This ratio is an indicator of the maximum brightness of
    lensing events in the extended source description (see
    Eq.~\ref{eq.DeltaFmax_FS}). {\it
      Right panel} Minimum distance that a source
    lens pair must have in order to cause an event as bright as
    WeCAPP-GL1 ($\DFmeas=10.15\times10^{-5}\E{Jy}$) as a function of the source characteristic (flux and
    radius) and lens mass. The {{\it light red}, {\it magenta}, {\it green}, {\it cyan}, {\it blue}, 
    {\it black}}  lines represent lenses with masses
    $0.1\Msun$,$0.5\Msun$,$1\Msun$,$5\Msun$,$10\Msun$, and $50\Msun$, respectively.}
  \label{fig.CMD_GL1}
\end{figure}

In Fig.~\ref{fig.CMD_GL1}b (middle panel) we show the 
  $F_0/R_*$-ratios for a bulge 12.6 Gyr SSP and 5 younger components
of the disk composite SP, where the flux $\F$ is given in
Jansky (Jy), the radius $\Rstar$ in units of the solar radius, and the
source is placed at 770 kpc.  The brightest bulge events, that can
occur, have an un-reddened color of about $\RI\approx 0.7$. 
For the ages above 500 Gyr the stars which can cause the brightest
lensing events are either very blue, or have a color of about
$\RI\approx 0.7$ (i.e. PMS stars present for the ages shown here). The
very red luminous stars with $\RI>1$ are ineffective sources for
bright lensing events since they are so large, implying a smaller
magnification at light curve maximum.
{The numerous (but fainter) turn-off stars of an 12.6 Gyr old
  stellar population would need a reddening of $E(R-I)\approx
  0.45\E{mag}$ to show the measured color of $\RI=0.83\E{mag}$.
  Because this translates into an extinction of $A_R\approx
  E(R-I)/(1-(\ext_I/\ext_V)/(\ext_R/\ext_V))=0.45/(1-0.482/0.748)=1.27\E{mag}$ 
  these stars can only produce $\RI=0.83$ events which are 13 times fainter than the brightest bulge stars.}

{The position of GL1 (3'6'' south, 2'40'' west of the M31 center) 
supports the assumption that the source star belongs to 
the bulge population.
Using the assumed extinction for the bulge of $\ext_R=0.36\E{mag}$ 
the intrinsic color of GL1 changes 
to $(R-I)_0=0.70$, which is also the source color for the brightest bulge events 
(see the peak of the flux-to-radius relation
for a 12.6 Gyr old bulge star in Fig.~\ref{fig.CMD_GL1}b).
Such a bulge star has an un-extincted absolute brightness  
of $\Mlum_R = -0.91\E{mag}$ and a radius of $\Rstar = 30\Rsun$. 
This is in good agreement to the (extinction corrected) R-band brightness of -1.00 mag  
obtained from the light curve fit (see Table~1 ).
}

In Fig.~\ref{fig.CMD_GL1}c (right panel) we show the minimum lens-source
distances $(\Dos-\Dol)$ as a function of brightness, radius and mass
$\F/(\Rstar/\Rsun)\times\sqrt{M/\Msun}$ using the measured flux excess of WeCAPP-GL1
$\DFmeas = 10\times 10.07^{-5}\E{Jy}$.
For bulge stars the largest ${\F 10^{-0.4\ext}/(\Rstar/\Rsun)}$ values
are of order $4\times 10^{-8}\E{Jy}$ (middle panel, red curve). 
{Now using this with the flux excess of GL1 combined with a bulge extinction of $A_R=0.36\E{mag}$, 
a $1\Msun$-lens needs a minimum source distance of about 8.1 kpc (green line in the right
panel of Fig.~\ref{fig.CMD_GL1})}, a separation where the product of
source and lens densities becomes small (Fig.~\ref{fig.CMD_GL1}a).

In Fig.~\ref{fig.analyse_GL1} we combined the Figs.~\ref{fig.CMD_GL1}a
and \ref{fig.CMD_GL1}c for the LOS of WeCAPP-GL1 and show the minimum
source-lens distances as a function of the source distance for different
self-lensing masses ($\Mlens=0.1,0.5,1,5,10,50\,\Msun$). { The two peaks
in the density contours arise from the maxima in the source density and lens density,
respectively.  It can easily seen that lowering $\Dos$ would
  slightly decrease the needed lens-source separations, but at the
  same time the source density decreases rapidly.}

\begin{figure}
  \centering
  \includegraphics[width=1.0\textwidth]{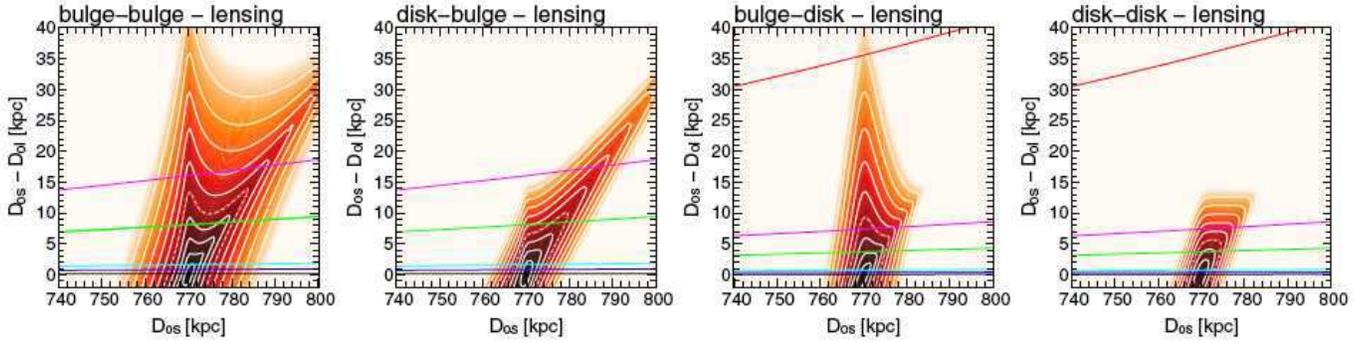} 
  \caption{Minimum lens-source distance for all self lensing
    configurations for the WeCAPP-GL1 event, using the intrinsic color
    of WeCAPP-GL1 of $(R-I)_0=0.70$ for bulge sources and $(R-I)_0=0.59$ for
    disk sources and Eq.~(\ref{eq.Dol_GL1}). The {{\it red}, {\it magenta}, {\it green}, 
    {\it cyan}, {\it blue}, {\it black}} lines represent lenses with masses
    $0.1\Msun$,$0.5\Msun$,$1\Msun$,$5\Msun$,$10\Msun$, and $50\Msun$, respectively. The
    contours show the product of the density of all source and lens
    stars (irrespective of color) as a function of the LOS distance to
    GL1 and the lens-source distance, $\rho_\M{source}
    (\vec\theta_\M{GL1},\Dos) \times
    \rho_\M{lens}(\vec\theta_\M{GL1},\Dol)$.  Contour levels are
    separated by factors of 10, the {\it dashed} contours mark a
    density of $\rho^2=10^{-4}$ ($\rho$ in units of solar masses per
    cubic parsec).  }
  \label{fig.analyse_GL1}
\end{figure}

For bulge sources at $770\E{kpc}$ in Fig.~\ref{fig.analyse_GL1} a,b
the minimum source-lens separation is $\Dos-\Dol = 74\E{kpc}$ for a
low mass lens with $\Mlens=0.1\Msun$, $16\E{kpc}$ for a $0.5\Msun$
lens, and $8.1\E{kpc}$ for $1\Msun$.
For masses lower than  $0.5\Msun$  
the large lens-source distance required to create the observed lensing event,
implies a very small event probability, because for large lens-source
LOS distances either the source or the lens density is
small.  
So, from just using the brightness of the event, lensing of bulge
stars $\le 0.5\Msun$ is very unlikely. 
For a bulge source more reasonable  lenses for self lensing are high mass
bulge remnants ($\Dos-\Dol = 0.16\E{kpc}$ for $50\Msun$) or very rare,
young, high mass disk stars ($0.82\E{kpc}$ for $10\Msun$, $1.6\E{kpc}$
for $5\Msun$).

Despite the lower disk density
at {the position of GL1} disk sources lensed by bulge remnants or high mass
disk stars seem more plausible (see the two right panels of
Figs.~\ref{fig.analyse_GL1}c and \ref{fig.analyse_GL1}d). 
{We used a reasonable disk extinction of $\ext_R=0.68\E{mag}$
with the source color and brightness derived from the 
light curve fit. These values are only consistent with a population 
older than $\ge 0.5$ Gyr (see blue marker in Fig.~\ref{fig.CMD}a).
Therefore for these disk sources 
the maximum  $\F/\Rstar$}-ratio (at  $(R-I)=0.59\E{mag}$) is obtained by 
stars with $\Mlum_R = -1.83\E{mag}$ and $\Rstar = 35.1\Rsun$.

For a disk source at $770\E{kpc}$ a lens with $1\Msun$ has to be 
separated by $3.7\E{kpc}$; for a lens with
$\Mlens = 5\Msun$ this separation becomes $ 0.75\E{kpc}$.  The
distances are smaller, since the disk sources are intrinsically
brighter, and thus a lower magnification is needed for the event, and
therefore, lenses can be closer to the source.

Of course, the event rate contribution is not just proportional to the
mass density product of all lenses and {\it all} sources (the product
of that is displayed as contours in Fig.~\ref{fig.analyse_GL1}) but
depends on the number density of those sources and lenses only that
can produce the observed event.  There, the $(M/L)$, and
other stellar population properties (fluxes of post-main sequence
stars, fraction of stars in a certain color interval) enter.

{However our qualitative result, that WeCAPP GL1 lenses below $0.5\Msun$ for bulge 
sources and below $0.1\Msun$ for disk sources are very unlikely, is consistent
to the quantitative result in Fig.~\ref{fig.pofM_fs_treatment_trueresult}.}

\section{Quantitative WeCAPP-GL1 analysis including finite stellar sizes }
\label{sec.quantitative}

In \S\ref{sec.pointsource} we treated stars as point sources, like
\cite{2003ApJ...599L..17R}.  
{We now abandon the point source approximation and use for each star
the radius based on the isochrones from \cite{2002A&A...391..195G}.}
We evaluate Eq.~\ref{eq.mass_probability}, using the stellar source
size distributions, and show the results in
Fig.~\ref{fig.pofM_fs_treatment_trueresult}.

\begin{figure}[h]
	\centering
  \includegraphics[width=0.9\textwidth]{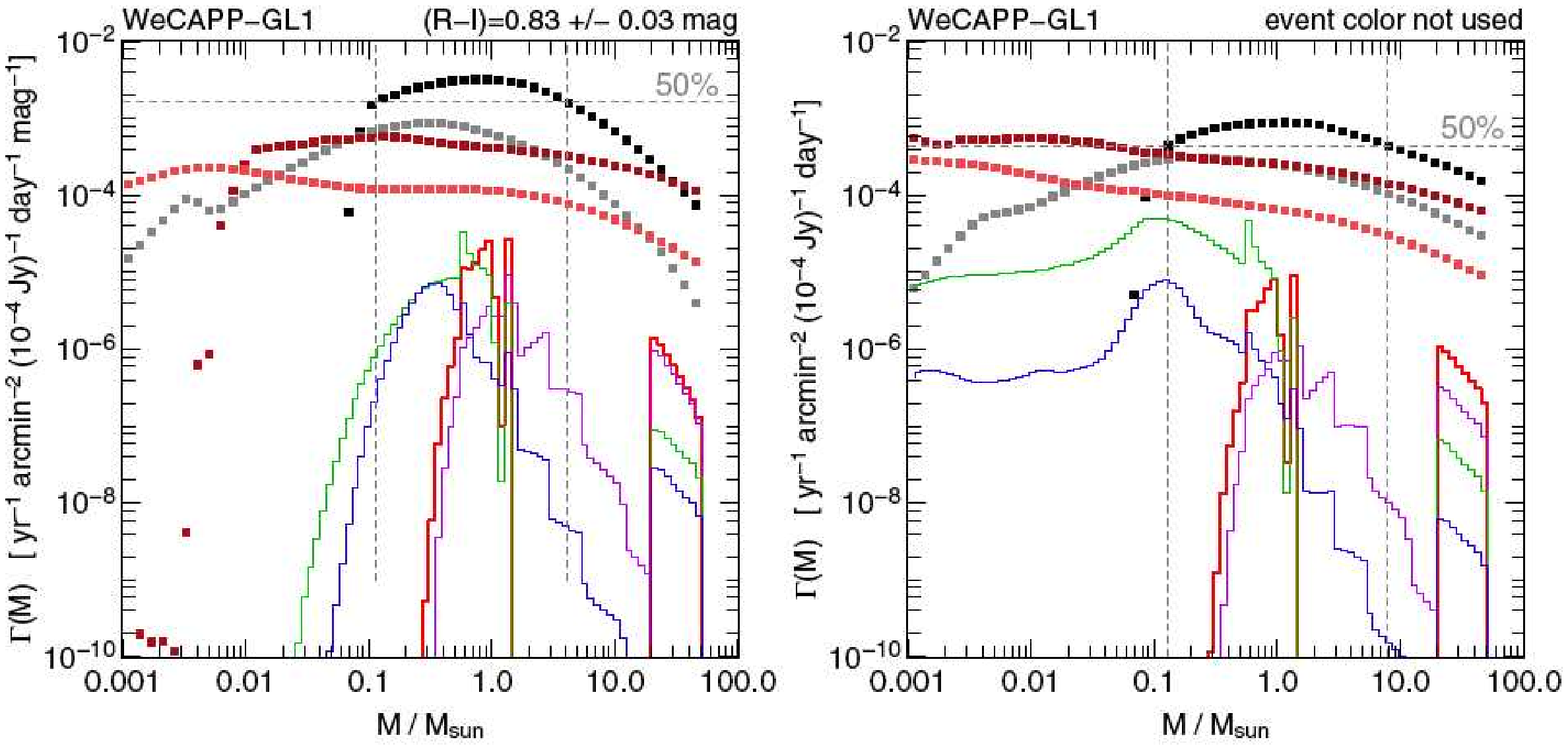}
	\caption{Event rate $\Gamma(\ml)$ (per area, per event brightness
    $\DF$, per event time scale $\tfwhm$, per color $\RI$) for
    WeCAPP-GL1 using the evolved MF including remnants.  The lens mass
    probability was determined allowing for {\bf finite stellar
      sizes}, using the measured $\tfmeas$, $\DFmeas$ and $\Colmeas$
    of the event [$(R-I)_0=0.59$ for disk sources, $(R-I)_0=0.70$ for bulge
    sources, with a Gaussian error of $0.1$], assuming a mean M31
    extinction of 0.19 and 0.51 for bulge and disk stars, and a MW
    extinction of { 0.17} (in the R-band). The {\it red}, {\it green},
    {\it blue} and {\it purple} curves show the lens mass distribution
    function for bulge--bulge, bulge--disk (bulge lens and disk
    source), disk--bulge and disk--disk lensing.  The information
    about halo lensing (M31-halo--bulge and M31-halo--disk data points
    are in {\it black} and {\it grey}, and MW-halo--bulge and
    MW-halo--disk in {\it brown} and {\it orange}) is displayed as
    points and not by a curve: we assume a {mono-mass-spectrum} for
    the halos, and each point represents a halo which is made of a
    given MACHO-mass to $100\%$.}
 	\label{fig.pofM_fs_treatment_trueresult} 
\end{figure}

Figure~\ref{fig.pofM_fs_treatment_trueresult} shows the halo lensing
and self lensing event rates (per flux excess and FWHM-time of the
event and per year and square arc minute) at the location of WeCAPP-GL1
using its brightness, time-scale and color and the extended stellar
sizes.  
{
Comparison to Fig.~\ref{improved_pointsource_p_of_M} shows, that self
lensing becomes dramatically suppressed; the vertical
scale changed by one order of magnitude relative to
Fig.~\ref{improved_pointsource_p_of_M}.
The halo-bulge lensing rate decreases by a factor of {6} relative
to the point-source formalism.  The self lensing rate however
decreases much more dramatically using the finite source description.
Relative to the most likely M31 halo-bulge lensing, self lensing
(bulge-bulge, disk-bulge, disk-disk and bulge-disk) is about 2 orders
of magnitude less likely, e.g. lensing of a bulge star by a
$0.8\Msun$ MACHO in M31 ({\it black points}) is {119} times more
likely as bulge-bulge ({\it red curve}) or bulge-disk ({\it green
  curve}) self lensing. The total halo-lensing contribution (summing
over M31 and MW halo-bulge and halo-disk lensing) is {63} times
more likely than all self lensing contributions assuming a 100\% MACHO
halo.
In addition, for each lensing configuration, the probability for small
lens masses is reduced. This can be explained as follows:
The large magnification needed for bright events can be obtained in
the point source approximation either by large Einstein radii
(efficient source-lens geometry or large lens masses) or events with
small impact parameter. For point sources, the magnification can
formally become infinitely large if the source passes the LOS to the
lens.  For extended sources the magnification saturates once the
source comes closer to the lens LOS than its projected source radius.
The source radius projected onto the lens plane is the larger, the
closer the lens-source pair, and therefore self lensing events are
most strongly suppressed in magnification and thus in flux excess at
the light curve maximum.  The only way to obtain bright events with
finite source sizes is to have spatially well separated source and
lens stars (increasing the Einstein radii and decreasing the projected
stellar sizes) or to have large lens masses (increase of Einstein
radii).
Therefore $0.1\Msun$ (for M31-halo--bulge lensing) and $0.01\Msun$
(M31-halo--disk lensing) MACHOs have too small Einstein radii to
achieve the required large magnification once finite source sizes are
taken into account. The suppression sets in for smaller lens masses in
the M31-halo--disk lensing case, because disk stars are brighter than
bulge stars.  
}

\begin{figure}
  \centering
  \includegraphics[width=1.0\textwidth]{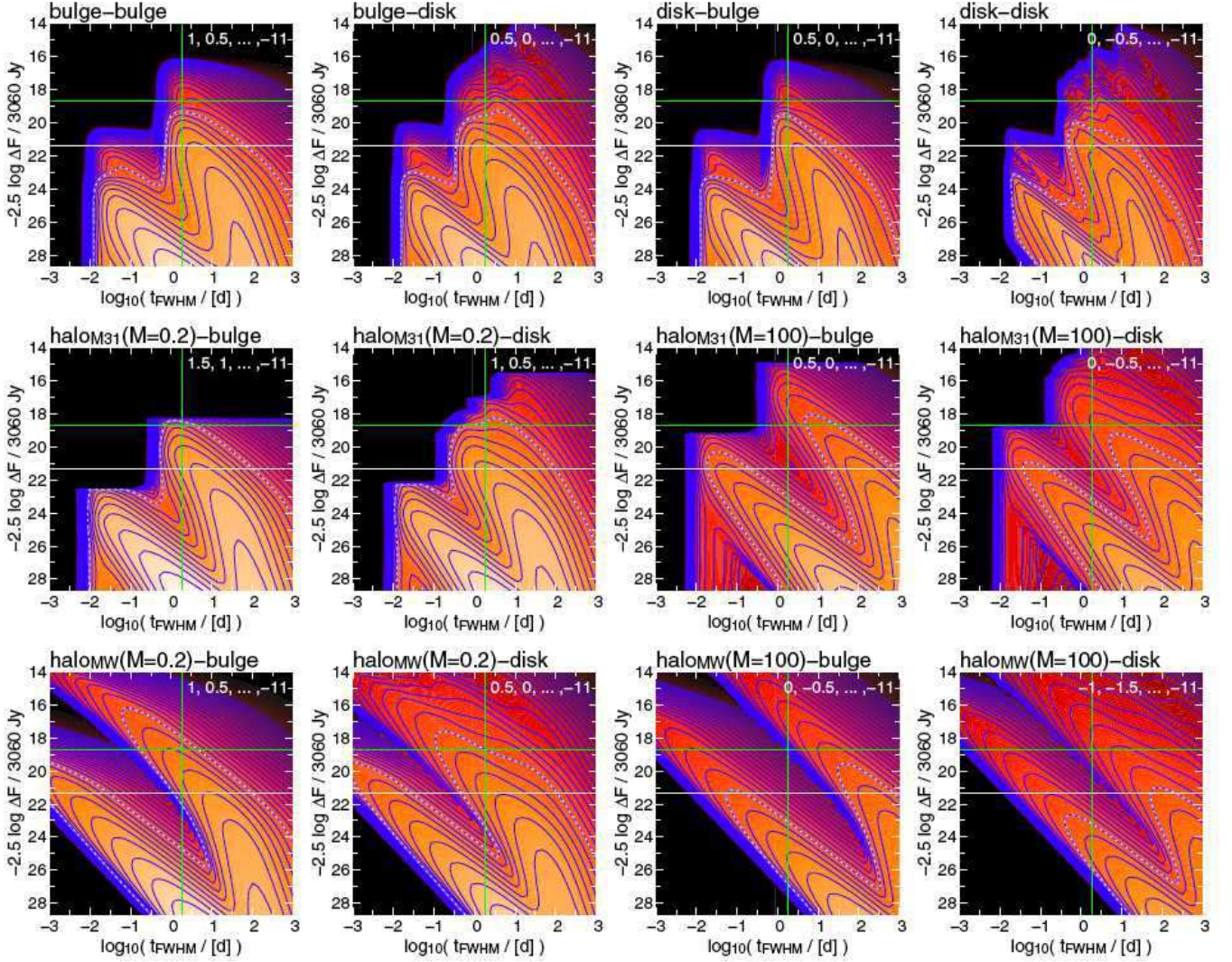}
  \caption{Flux excess and time scale distribution of events {\bf
      with} WeCAPP-GL1 {\bf color} ($(R-I)_0=0.59$ for disk sources,
    $(R-I)_0=0.70$ for bulge sources, without using its error) and its
    location. We make use of {\bf finite stellar sizes} in
    Eq.~\ref{eq.eventrate_obs_5dim_log} but display only events which
    do {\bf not show any finite source signatures (no fss)} in their
    light curves (left part of brackets in rhs part of
    Eq.~\ref{eq.eventrate}). The two further observables of WeCAPP-GL1
    (flux excess and event time scale obtained from a point source
    lensing light curve fit) are marked in {\it green}.  The estimate
    for the WeCAPP $6\sigma$ detection limit (see
    \cite{2006ApJS..163..225R}, Table 1) at the position of GL1 is
    shown as {\it white line}.  The contours differ by $10^{-0.5}$,
    and the dashed contour is that of $10^{-3}$ events per year, per
    square arc minute at location of WeCAPP-GL1, per color (in
    magnitudes), per logarithmic timescale (time scale in days) and
    per flux-excess (converted to magnitudes) (see
    footnotes~\ref{fn.mag_flux} and \ref{fn.conversion}).}
\label{fig.events_tf_DF_nfs} 
\end{figure}

\begin{figure}
  \centering
  \includegraphics[width=1.0\textwidth]{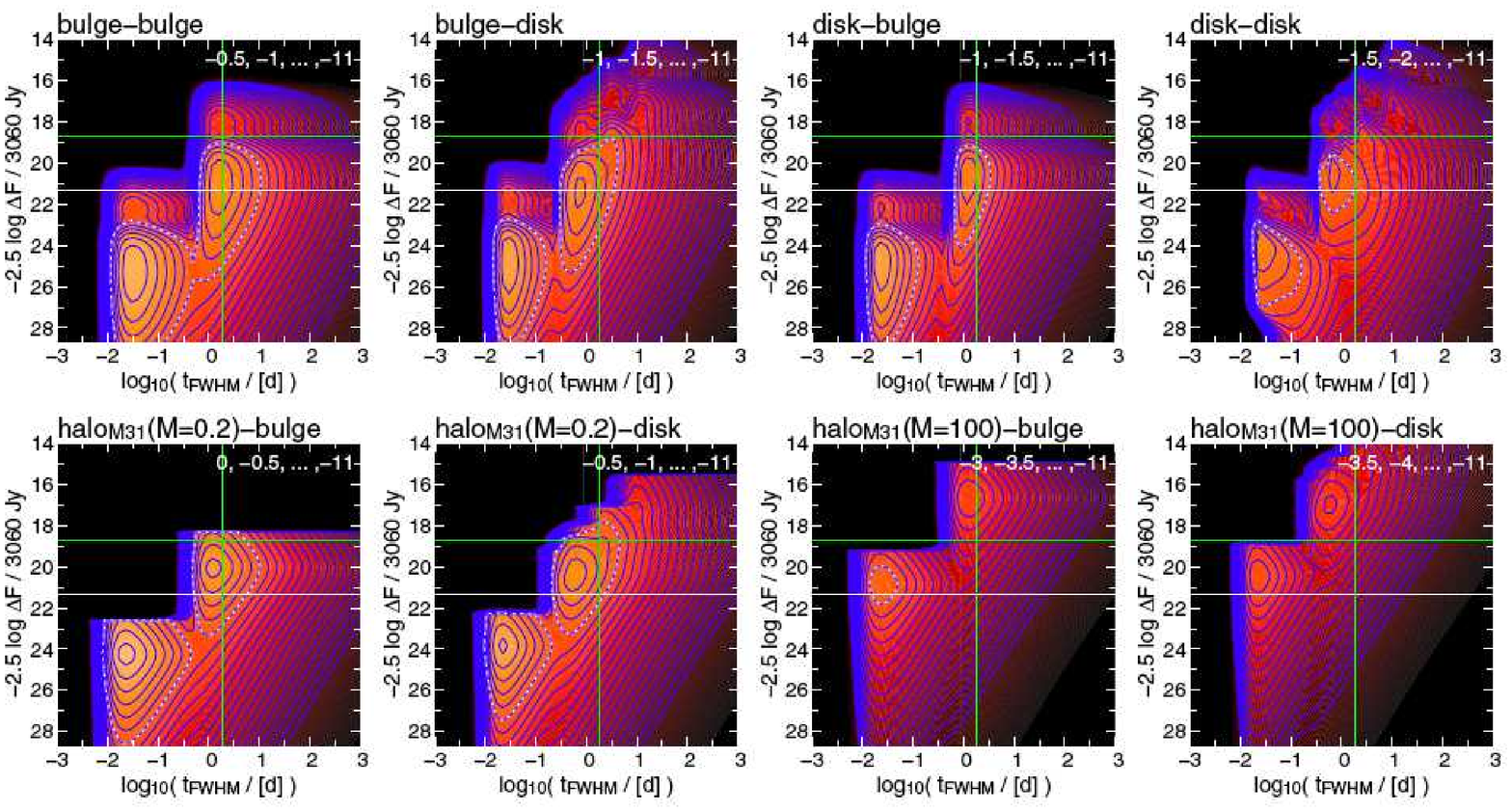}
  \caption{Flux excess and time scale distribution of events {\bf
      with} WeCAPP-GL1 {\bf color} [$(R-I)_0=0.59$ for disk sources,
    $(R-I)_0=0.70$ for bulge sources, without using its error] and its
    location. We account for {\bf finite stellar sizes}
    in Eq.~\ref{eq.eventrate_obs_5dim_log} but add only events
    (right part of brackets in rhs part of Eq.~\ref{eq.eventrate})
    which do {\bf show finite source signatures (with fss)}.  The two
    further observables of WeCAPP-GL1 (flux excess and event time
    scale obtained from a point source lensing light curve fit) are
    marked in {\it green}. { The estimate for the WeCAPP $6\sigma$
      detection limit (see \cite{2006ApJS..163..225R}, Table 1) at the
      position of GL1 is shown as {\it white line}.}  The contours
    differ by $10^{-0.5}$, and the dashed contour is that of $10^{-3}$
    events per year, per square arc minute at location of WeCAPP-GL1,
    per color (in magnitudes), per logarithmic timescale (time
    scale in days) and per flux-excess (converted to magnitudes) (see
    footnotes~\ref{fn.mag_flux} and \ref{fn.conversion}).}
	\label{fig.events_tf_DF_wfs}
\end{figure}

To illustrate the differences between point source and extended source
sizes approximation in more detail, we evaluate
Eq.~(\ref{eq.eventrate_obs_5dim_log}) and separate the events into
those which do not show any finite source signatures in their light
curves (no fss) and those which show finite source signatures (with
fss) and display the corresponding results in
Figs.~\ref{fig.events_tf_DF_nfs} and \ref{fig.events_tf_DF_wfs}.
\footnote{To convert to the same units as in
  Fig.~\ref{fig.pofM_fs_treatment_trueresult} use
  Eq.~\ref{eq.eventrate_obs_5dim_log}. E.g. in Figs.~\ref{fig.events_tf_DF_nfs}
  and~\ref{fig.events_tf_DF_wfs} the WeCAPP-GL1
  measurements cross the contour level of $\M{halo}_\M{M31}(\ml$=$0.2\Msun)$-bulge
  lensing at -2.30 and -1.86, respectively. This
  converts to $\Gamma / [\M{yr}^{-1}\E{arcmin}^{-2}\E{mag}^{-1}] = (10^{-2.30}+10^{-1.86})/(0.4 \ln(10)^2\cdot1.83\E{day}\cdot10.07\times 10^{-5}\E{Jy}) \times  10^{-4} = 4.8 \times 10^{-3}\;(10^{-4}\E{Jy})^{-1}\E{day}^{-1}$.\label{fn.conversion}}
To understand the differences to Fig.~\ref{fig.events_tf_DF_all} it is
useful to recall the relations for event time scale $\tfwhm$ and
magnification of events at light curve maximum $\A$ in the high
magnification approximation [for carrying out the integrals we used
the exact relations described in \cite{2006ApJS..163..225R}]:
\begin{equation}
  \left\{
    \begin{array}{ll}
      t^\M{no\ fss}_\M{FWHM} (b)  
      \equiv \tfwhm(b)
      \propto \tE u_{0} = \tE {b \over \RE} = \vt ^{-1} \, b
      & ,\quad \quad  b \ge \bfs 
      \\
      t^\M{with\ fss}_\M{FWHM} (b) 
      \equiv  \tfsFWHM(b)
      \approx t_\M{FWHM}(b) {1\over \sqrt 3 } \sqrt{ (2\bfs/b)^2 -1  }
      > \tfwhm (b) & ,\quad \quad  b < \bfs
    \end{array}
  \right.
\end{equation}
\begin{equation}
  \left\{
    \begin{array}{ll}
      A_0^\M{no\ fss}(b) \equiv  \A(b) \approx  \RE/ b
      & ,\quad \quad  b \ge \bfs 
      \\
      A_0^\M{with\ fss} (b) \equiv  \A(\bfs) \approx
      \RE/ \bfs
      & ,\quad \quad  b < \bfs
    \end{array}
  \right.
\end{equation}
with $\bfs \approx \Rstar \Dol / ( 2 \Dos )$. 
In the point source approximation, source lens configurations
with small transverse velocity and small impact parameter can
contribute to the event rate at given $\tfwhm$; in the finite source
treatment, transverse velocities have to be larger than a limit $\vtfs
\propto \bfs /\tfwhm$ to make events without finite source signatures
(no fss); in the other case, $\vt<\vtfs$, light curves will show
finite source signatures (with fss) (i.e. show a saturated
magnification, and an increased event time scale). Therefore,
Fig.~\ref{fig.events_tf_DF_wfs} can be constructed out of
Fig.~\ref{fig.events_tf_DF_all}, by {moving} the event rate
contributions due to events with $\vt < \vtfs$ to events with finite
source signatures, after accounting for their increased time scale,
and their decreased brightness.
By comparing the panels in the last {row} in figures
Figs.~\ref{fig.events_tf_DF_all} and \ref{fig.events_tf_DF_nfs} one can
see that finite source effects are almost unimportant for Milky Way
MACHO-events with lens masses between $0.2\Msun$ and $100\Msun$: for
both MACHO masses, the event rate and the characteristics of the
events with time scales larger than 1 d (which are observable with
present experiments) are nearly the same in the point source
approximation and in the extended source size treatment.  The
remaining panels in Fig.~\ref{fig.events_tf_DF_nfs} show, that finite
source effects are { less} important for heavy M31-MACHOs, and are
relevant for low mass M31-MACHOs, like $0.2\Msun$ (second {row} 
sub panels), and that they are extremely important for the correct
interpretation of the self lensing contribution (first {row}
sub panels): an event like WeCAPP-GL1 can hardly be caused by MACHOs
with masses much smaller than $0.2\Msun$, otherwise the flux-excess
cutoff would be lower than the event brightness, and WeCAPP-GL1 is
unlikely to be caused by self lensing. Among all self lensing
scenarios (Fig.~\ref{fig.events_tf_DF_nfs}, first row) the brightness of
WeCAPP-GL1 is most easily to achieve by bulge lenses
(Fig.~\ref{fig.events_tf_DF_nfs}, {\it first two panels of the first
  {row}}), because in that case source and lens stars are relatively
separated (more than for disk lenses).

\begin{table*}[h]
  \scriptsize
  \setlength{\tabcolsep}{1mm}
  \centering
{
  \begin{tabular}{l|l|l|l|l}
type              & $\DF_R$         &  $\magR(\DF_R)$ & event rate  & mean time between events \\
                  & [$10^{-5}$ Jy ] &  [mag]          &   [ev/yr]   & [yr] \\
  \hline  \hline

self lensing      & $\ge   10$      &  $\le 18.7$     &      0.0101 & 99.2    \\
''                & $\ge    8$      &  $\le 19.0$     &      0.0204 & 49.0   \\
''                & $\ge  0.9$      &  $\le 21.3$     &       6.47   & 0.155  \\
  \hline
halo lensing with   
20\% $1\Msun$     & $\ge   10$      &  $\le 18.7$     &      0.0541  & 18.5   \\
''                & $\ge    8$      &  $\le 19.0$     &      0.0969  & 10.3   \\
''                & $\ge  0.9$      &  $\le 21.3$     &      2.21    & 0.45 \\
 \end{tabular}								 
 \caption{Expected rates for short time scale (FWHM-timescales between 1 
   and 3 days) lensing events in the entire 17.2'$\times$17.2' WeCAPP,
   assuming 100\% detection efficiency.} 
}
\end{table*} 

{
In Table~3 we have evaluated the number of events we expect
to find with time scales between 1 and 3 days in the entire 17.2'
$\times$ 17.2' field monitored by WeCAPP (assuming 100\% detection
efficiency). Short self lensing events with a peak
magnitude\footnote{We define the function $\magR(x)$ to replace the
  transformation from fluxes to magnitudes $\magR(x) \equiv
  -2.5\log_{10}(x/F_\M{Vega,R})$ with $F_\M{Vega,R}=3060\E{Jy}$.  The
  inverse is defined as $\fluxR(x) \equiv F_\M{Vega,R} 10^{-0.4
    x}$.\label{fn.mag_flux}} larger than $19.0\E{mag}$ (as lower limit for
WeCAPP-GL1, see \S\ref{sec.qualitative}) in the should take place only
every 49 years. For reasonable halo mass fraction of 20\% consisting
of $1\Msun$-MACHOs every 10 yrs (short) events above a $19.0\E{mag}$
threshold could be observed.  These values decrease very rapidly
lowering the threshold to the most likely peak brightness for GL1 of
$18.7\E{mag}$: The mean time between events for selflensing is 99 yr
whereas for halo lensing 18 yrs.
While for the short, fainter events ($\le 21.1\E{mag}$) self-lensing
would dominate by a factor of 3.
Note that these values represent the theoretical event rates in the
WeCAPP field without detection bias. In other words only if a survey
is complete to a certain threshold it should be possible to measure
these numbers of events.  For real microlensing surveys the detection
efficiency for bright events will be much higher than for faint
events, and therefore the ratio of bright vs. short events would be
different.
If we assume to have the same detection efficiency for halo and
selflensing (which in principle can differ according to there
intrinsic distribution) the difference between halo-lensing and self
lensing favors the assumption that WeCAPP-GL1 ({\it green line}) was
caused by a halo lens. }

{To summarize, an event as bright as WeCAPP-GL1 is extremely
  unlikely to be a self lensing event. The most likely self lensing
  scenario is bulge-disk self lensing. MACHO-lenses above 0.2 solar
  masses are expected to cause the bright events much easier.}

\section{Interpreting WeCAPP-GL1: impact of  extinction}
\label{sec.extinction}

\cite{2006ApJS..163..225R}, which assumes an {on average}
M31-extinction for all disk stars in the WeCAPP field of $0.51\E{mag}$ and
for all bulge stars of $0.19\E{mag}$ in the R-band, independent of the
angular position of the event, and independent of the LOS distance to
the source.  The MW-extinction is set to $0.17\E{mag}$ (R-band). 
The result
for { this} M31 standard-dust { model} has been described in
Fig.~\ref{fig.pofM_fs_treatment_trueresult} already.

To see to which extent dust can change the interpretation of lensing
events, we set the M31-extinction (not the MW-extinction) along the
LOS to WeCAPP-GL1 to zero.  We do not assume that the mean
M31-extinction of bulge or disk stars goes to zero as
well\footnote{This would imply a decrease of the disk and bulge number
  density and thus of the event rates for disk source stars by {
    $37\%$} and for bulge source stars by { $16\%$} and thus would
  decrease the lensing rate of disk sources relative to bulge sources},
but that by {accident} just this one sight line to the source is not
extincted by M31-dust at all; the extinction then equals that of the
MW of about $0.17\E{mag}$.  
The results are shown in the {left-most column of} panels in
Fig.~\ref{fig.final_eventrates_1}.  
With no M31-dust
along the LOS to WeCAPP-GL1 { the event would} become
intrinsically fainter (relative to the extinction case and giving the
observed fluxes) by $0.19\E{mag}$ and $0.5\E{mag}1$ magnitudes in the
R-band for bulge and disk sources.  This reduces the necessary
magnification and makes self lensing more likely.  At the same time
the lensed sources are expected to be intrinsically redder, which
changes the stellar types of stars and thus its absolute brightness ({\it lower panels}).  
One can derive the increase of the event rates
for an event like WeCAPP-GL1 also by shifting the contours in
Figs.~\ref{fig.events_tf_DF_nfs} and \ref{fig.events_tf_DF_wfs} by
$0.19\E{mag}$ and $0.51\E{mag}$ along the $\DF$-direction for lensing
of bulge sources and disk sources respectively.  The ratio for the
event rates of all self-lensing to all halo-lensing configurations is
still of the order of 1:100 for full MACHO halos and MACHO masses in
the range of 0.1 to 1 solar masses. So even in the implausible case of
no M31-dust along the LOS to the source of WeCAPP-GL1, a $1\%$ MACHO
fraction of the M31 and MW halos already provides the same lensing
rates as self lensing does.

As alternative values for the total (MW+M31) line of sight extinctions
we also use values of $0.7\E{mag}$ and $1.05\E{mag}$ ({``strong
extinction''}) and show the results in the third and fourth columns of
Fig.~\ref{fig.final_eventrates_1}.  In this case lensing rates are
suppressed and the most likely lens masses are shifted to higher
masses. This can be understood since now only higher mass lenses are
able to provide the high flux excess found for WeCAPP-GL1.

\begin{figure}[h]
	\begin{center}
     \includegraphics[width=1.0\textwidth]{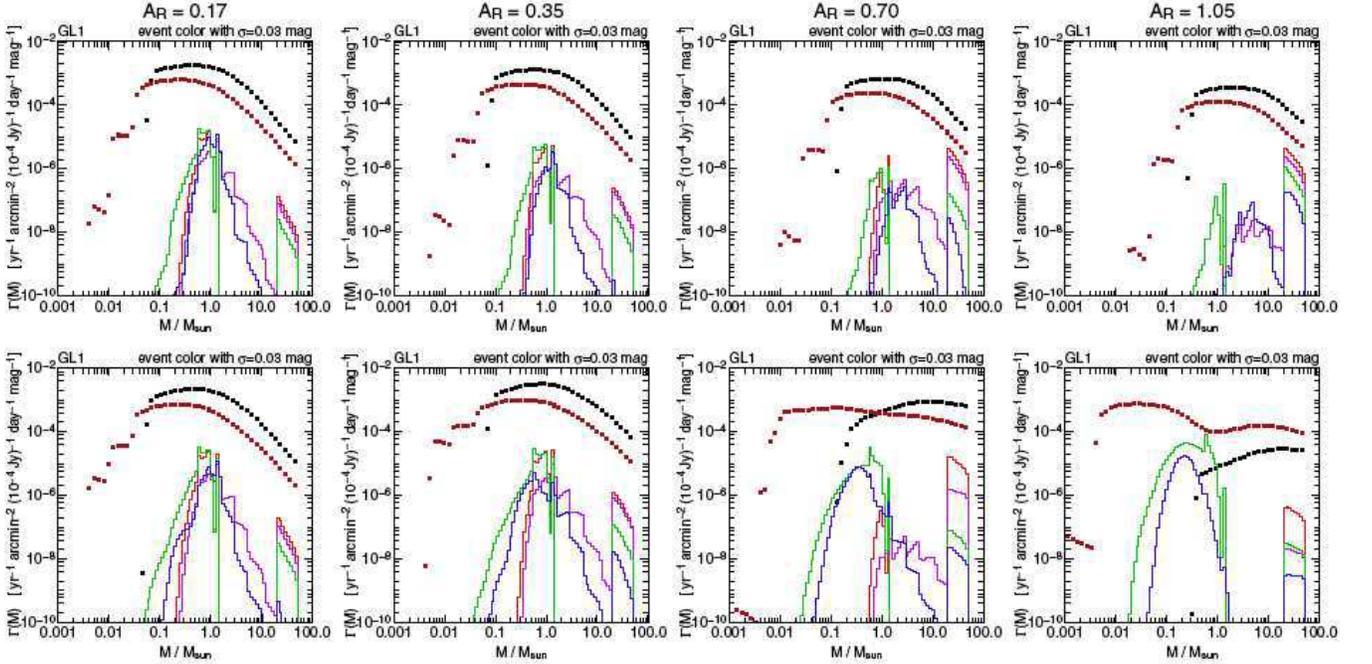}
\end{center}
\caption{Event rate $\Gamma(\ml)$ (per area, per event brightness
  $\DF$, per event time scale $\tfwhm$, per color $\RI$) for
  WeCAPP-GL1.  {\it upper panels:}
  without changing the color by the assumed extinction [{$(R-I)_0=0.59$ for disk sources, $(R-I)_0=0.70$ for bulge
  sources, with a Gaussian error of $0.03$}], {\it lower
    panels:} taking into account the color shift due to the assumed
  extinction.  The {\it red}, {\it green}, {\it blue} and {\it purple}
  curves show the event rate for bulge-bulge, bulge-disk (bulge lens
  and disk source), disk-bulge and disk-disk lensing. The information
  about halo lensing (M31-halo--bulge and M31-halo--disk data points
  are in {\it black} and {\it grey}, and MW-halo-bulge and
  MW-halo-disk in {\it brown} and orange) is displayed as points and not by
  a curve: we assume a {mono-mass-spectrum} for the halos, and each
  point represents a halo which is made of a given MACHO-mass to
  100\%.
  The results have been obtained with an average dust extinction of
  total bulge and disk population of $\ext_R=0.36\E{mag}$ and
  $\ext_R=0.68\E{mag}$.  These values are expressed in the two middle
  panels as 0.35 and 0.7. The results for the minimum line of sight
  extinction of 0.17 (set by the MW) and for a high extinction of 1.05
  are shown to the left and right of the same row. The 4 panels in
  the first row show results analogous to the 2$^{nd}$ row, with the
  difference, that the de-reddening of the source star has been
  neglected: this simplification would have a minor impact on the
  interpretation for the stellar populations assumed here.  }
  \label{fig.final_eventrates_1}
\end{figure}

In the second row of Fig.~\ref{fig.final_eventrates_1} we show how
the extinction influences the color of the event and therefore the
type of source stars which were most likely lensed.  Whereas in the
upper panels the event rates are shown, if one accounts for the
dimming but not for the reddening by dust, in the lower panels we take
into account the reddening with the assumed extinction.  The effect is
lowest for low extinction; for high extinctions the predicted masses
are strongly changed.  The assumed shifts in color $\RI$ for event
extinctions of 0.17, 0.35, 0.70, 1.05 mag are -0.06, -0.12, -0.25,
-0.37, respectively, which means that the {\bf intrinsic} colors of
GL1 are {0.77, 0.71, 0.58, 0.46}.  Comparing these colors with the
CMD in Fig.~\ref{fig.CMD}, shows that the intrinsic brightness can
drastically change for slight color differences: for the 12.6 Gyr component from
-1.0 to +3.7 mag, for the 2.5 Gyr disk component from -1.8 to +2.3,
  whereas the 20 Myr disk component slightly brightens from -6.2 to -6.9
  mag. This can explain the extreme differences in the lens mass
distributions and underlines the importance of accurate color
measurements and precise extinction estimates. Although the color
plays an important role in the correct mass interpretation, the ratio
between halo- and self-lensing is almost independent of the assumed
extinction.  For all extinction assumptions WeCAPP-GL1 is more likely
caused by a halo lens than by a stellar lens.
{Only for the de-reddened, high extinction case ({\it last panel,
    $2^{nd}$ row}) the halo to self-lensing ratio approaches a factor
  of {10}.  }

\clearpage

\section{The impact of a metal poorer  disk}
\label{sec.youngdisk}

The maximum flux excess $\DF$ of an event depends on the flux and
radius of the source star. Therefore, the question arises, how
sensitive the interpretation, that WeCAPP-GL1 can hardly be achieved
by self lensing, depends on the assumed {metalicity} of the stellar population.
The luminosity and sizes of post-main-sequence stars change {slightly}, 
if the metalicity of the composite stellar population is altered. 

We therefore have modeled a metal poorer disk \citep{2002MNRAS.331..293W} by changing 
the metalicities to $Z=0.008$ (see Girardi's {\tt isoc\_z008.dat}).  
The change of metalicity has several competing effects:

\begin{enumerate}
\renewcommand{\labelenumi}{\roman{enumi})}
\item a metal poorer population is brighter
\item for a metal poorer population, smaller MACHOs masses are
  allowed, since brighter stars need smaller magnification, which can
  be produced also by smaller masses without finite-source-size
  saturation of the magnification. 
\item the lens properties are nearly not affected by the luminosity
  evolution, since reducing the metalicity mostly changes the
  properties for the lensed stars (sources), and mass evolution takes
  place only for very few stars.
\item {a stellar population with lower metalicity (changed from solar
  by a factor of 2) contains PMS star brighter by a factor of roughly
  $2$ ($\approx 0.75\E{mag}$). Their radii ($\Rstar \propto
  \sqrt{F_0}$) are larger by a factor of roughly $\sqrt{2}=1.4$.
  Since (see Eq.~(\ref{eq.DeltaFmax_FS})) $\DFmax \propto F_0/\Rstar
  \propto \sqrt{F_0} \propto \Rstar$ for events with the same color,
  one expects the flux excess of the brightest events to increase by a
  factor of 1.4, or $0.38\E{mag}$ if the stars become metal poorer.
  This would make the contours in Figs.~\ref{fig.events_tf_DF_nfs} and
  \ref{fig.events_tf_DF_wfs} to shift by $0.38\E{mag}$ in vertical
  direction and push the flux-excess limit due to finite source
  effects by the same amount.}
\item However, if stars are brighter, fewer of them are needed to
  account for the total observed light, which reduces the overall
  lensing rate, e.g. the $(M/L)$ of the disk population drops from 1.2
  to 1.0 in the R-band if the metalicity of the population is changed
  from solar to 0.008.
\end{enumerate}

All these points explain the relative small differences in the event
rates by changing the metalicity of stellar population.  {However
  the lower part of mass probability distribution for a measured event
  can slightly shift to lower masses, as the lower amplifications can
  be produced by lower masses.}

\section{Outlook: Statistical interpretation - 
Halo vs. self lensing from brightness distributions of events}
\label{sec.statistics}

We now show that the brightness and time scale distribution of
  events can be used to discriminate halo and self-lensing.

Figure~\ref{fig.distribution_of_events} ({\it left panels}) compares
the event rates as a function of event brightness for self lensing and
halo lensing events within rings around the M31 center.  These rings
have a thickness of 1 arcmin and outer radii between 1 arcmin and 12
arcmin (blue/red for self lensing, green/black for halo lensing). The
event rates were calculated ignoring detection efficiency factors,
assuming a 12.6 Gyr bulge and a composite disk population, and the
mean extinctions for the bulge and the disk sources. We consider only
events longer than 1 day ($\tfwhm$). For self lensing, the event rate
drops steeper with event brightness within central rings 
than within outer rings. {This is caused because in the inner rings 
self lensing of bright events is suppressed by the closer LOS distances 
of source-lens pairs. For the outer rings the mean distances between
disk-bulge pairs are larger and the lensing rate is increased.}  
For halo lensing there is no such effect.
The brightness cut-off of the event-rate--event-brightness relation is
determined mostly by stellar population properties (self lensing) and
by the MACHO mass (halo lensing), as can be seen by 
comparing the {left panels} in Fig.~\ref{fig.distribution_of_events} which were obtained
for $0.1\Msun$, $1.0\Msun$, and $10\Msun$ MACHOs {from top to bottom}.  This demonstrates
that the brightness distribution of {bright} events could itself
discriminate halo and self lensing if a sufficient large number of
events is available. The two lines represent the WeCAPP-GL1 flux
excess (fit and highest data point).

\begin{figure}[h]
	\begin{center}
    \includegraphics[width=1.0\textwidth]{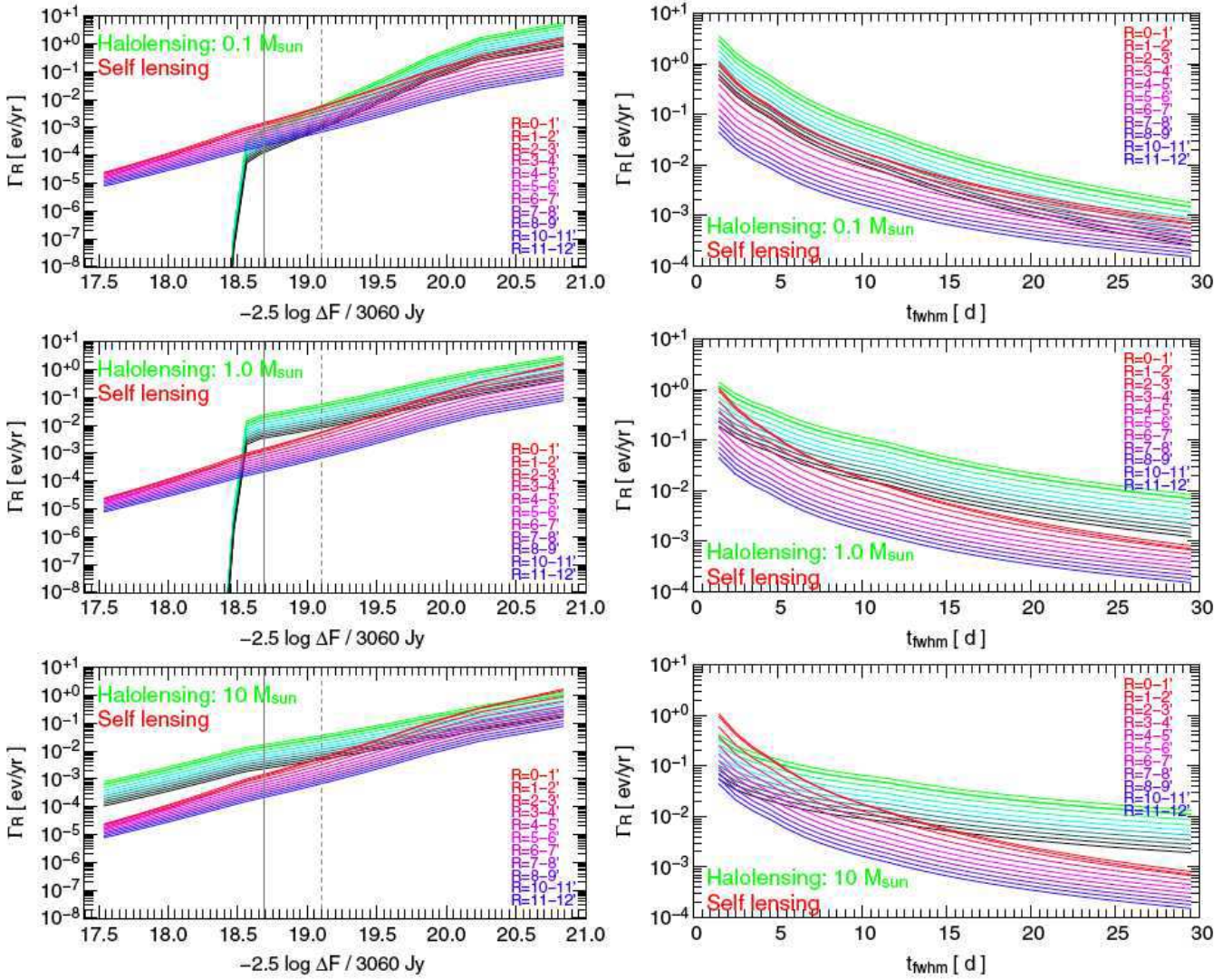}
  \end{center}
  \caption{{\it Left panels:} Brightness distribution of events (with
    $\tfwhm>1d$) within rings of 1 arcmin around the M31 center. Outer
    radii are chosen between 1 arcmin ({\it red} for self lensing,
    {\it black} for M31 halo lensing) and 12 arcmin ({\it blue} for self
    lensing, {\it green} for halo lensing). The two vertical lines
    ({\it grey}) represent the GL1 flux excess (fit and highest data
    point).  {\it Right panels:} Time scale distribution of events
    (brighter than $\DF>10^{-5}\E{Jy}$ corresponding to an excess
    magnitude of $21.2$) within rings of 1 arcmin around the M31
    center. Outer radii are chosen between 1 arcmin ({\it red} for
    self lensing, {\it black} for M31 halo lensing) and 12 arcmin ({\it
      blue} for self lensing, {\it green} for halo lensing)}
  \label{fig.distribution_of_events}
\end{figure}

The distribution of event time scales (only events brighter than
$10^{-5}\E{Jy}$ in R are considered) is fairly similar for self
lensing and MACHO events (in slope) if the MACHO mass is $0.1\Msun$
(this can also been seen from Fig.~\ref{fig.events_tf_DF_nfs} which shows
for the location of GL1 that the event characteristics [flux excess
and time scale] are most similar for self lensing and halo lensing if
the halo population has a mass of $0.2\Msun$). However, if MACHO
masses are larger, events will have longer timescales than predicted
from self lensing (see Fig.~\ref{fig.distribution_of_events}, right panels).
\\
The right panels in the $2^{nd}$ and $3^{rd}$ row of
Fig.~\ref{fig.distribution_of_events} show the time-scale
distribution for $1\Msun$ and $10\Msun$ MACHOs, respectively. 
Because of the clear difference between selflensing and halolensing the
time scale-distribution is a good discriminator if the MACHO mass is
larger than $1.0\Msun$.

The sharp cut-off in the brightness distribution of halo-lensing events is
caused by the fact that the MACHO mass function used is a $delta$-function.
Since the different lensing masses produce a brightness distribution
which has quite a different gradient with respect to selflensing, it is 
conceivable that there exists a specific halo-mass function where
the brightness distributions look similar to self-lensing.
Therefore, one probably can find a halo mass function that can mimic self
lensing as long as only the brightness distribution of events is
considered. If one however combines the brightness distribution with
the distribution of the time scales and locations of the events, the
halo-self-lensing degeneracy can be broken.

\clearpage

\section{Discussion and Conclusion}
\label{conclusions}

We showed that accounting for extended sources can
dramatically change the lensing and self lensing rates for events as
bright as WeCAPP-GL1. The reason is that magnification saturates, and
that the maximal brightness of an event for a given source size and
source flux depends only on the source and lens distances and lens
mass (not on the impact parameter): Very bright events thus require a
minimum source lens distance, which makes halo lensing more likely
relative to self lensing. For WeCAPP-GL1 the inclusion of the finite
stellar sizes makes the ratio of self lensing to halo lensing by about
a factor of 8 less likely compared to the point source approximation.
This result was obtained based on simple stellar population models of
the bulge and disk and on a simple description of the dust extinction
and is almost independent of MACHO mass as long as that is in a range
of $0.1-3.5\Msun$.

Likewise if one uses position, FWHM time scale, flux excess and
color of WeCAPP-GL1, self-lensing is even 13 times less likely than
lensing by a MACHO, if 1 solar mass MACHOs contribute only 20\% to the
total halo mass in this mass range.

Assuming a 100\% detection efficiency we expect a self lensing
  event with with FWHM time-scales between 1 and 3 days and a peak
  flux excess brighter than $19.0\E{mag}$ in the {\it entire} WeCAPP survey
  field only once every 49 yrs (99 yrs for $\magR(\DF)<18.7\E{mag}$).
On the other hand if 20\% of the halos of M31 and the MW are made
  of 1.0 solar mass MACHOs then an event like WeCAPP-GL1 would occur
  every 10 years (every 18 yrs for a $18.7\E{mag}$ threshold).

This implies that a small fraction of baryonic MACHOs like (brown
dwarfs, neutron stars, stellar black holes, cool white dwarfs) in the
M31 halo is sufficient to make WeCAPP-GL1 more likely to be
a halo-lens event.

We have also shown that different values for the extinction of the
WeCAPP-GL1 event do only slightly change the halo-lensing -- self
lensing ratio, but more strongly change the most likely lensing
masses. This is because the major impact of the extinction 
is that of a bluer intrinsic source color and therefore implies a 
change of stellar source size and brightness.

We emphasize that any interpretation for high difference flux events
and their predicted rates has to account for finite-source effects.
{Further published examples are
\cite{1999A&A...344L..49A,2001ApJ...553L.137A,2003A&A...405...15P}.}
Therefore the brightness distribution of events in general is a good
discriminator between self and halo lensing. The time-scale
distribution is a good discriminator if the MACHO mass is larger than
$1.0\Msun$.

Our model reasonably describes the $(M/L)$ for the stellar population
in M31 and we are confident that the mass density and velocity
distribution of the stars is fairly well modeled.  With these
ingredients one cannot obtain an event rate high enough to make a
detection of WeCAPP-GL1 likely within our survey interval. Therefore,
our analysis supports the existence of a (small baryonic or
non-baryonic) MACHO component in the halos of M31 or MW.

{In this paper we have ignored detection efficiencies.  We were
  allowed to do so, because, for a single event, the probability {\it
    ratio} of halo lensing to self-lensing is independent of detection
  efficiencies. For WeCAPP-GL1, halo lensing (assuming a 20\% MACHO
  halo) is 5 times more likely than self-lensing.  We admit that this
  factor of 5 is not large enough to conclude that self-lensing is
  definitely excluded. However, we point out, that 'WeCAPP-GL1'-like
  self-lensing events are very rare.  Put in another way, the fact
  that this event was observed can be more easily understood, if there
  is a, e.g., 20\% compact dark matter fraction in the halos of M31
  and the MW.
  The analysis of more bright events, and the comparison of the event
  rates for bright and faint events will give more insight, as well as
  the use of their time scale distribution. This however requires the
  knowledge of the detection efficiency (as a function of brightness,
  timescale, location and color) of the events, and will be subject of
  a further paper.
  Our results provide a strong motivation to search for the brightest
  short time scale events towards M31 in wide field surveys (like
  Pan-STARRS 1).  
}

\appendix

\section{M31-model}
\label{model}

Assuming an M31-distance of 770 kpc we converted our light model
(described in \cite{2006ApJS..163..225R}) into mass using the
theoretical bulge and disk R-band $(M/L)_R$ of $3.3$ and
$1.2$ derived using their theoretical stellar populations.
The colors extracted from these stellar populations are for
the bulge $(B-R)_0 = 1.63$ [$(B-R)_\M{meas} = 1.91$], $(V-R)_0 =
0.60$ [$(V-R)_\M{meas} = 0.72$] and for the disk $(B-R)_0 =
0.81$ [$(B-R)_\M{meas} = 1.33$], $(V-R)_0 = 0.37$
[$(V-R)_\M{meas} = 0.60$].
The analytical mass density models are
\begin{equation}
  \rho_\M{bulge}(x,y,z) := 44\,100 \Msun/\M{pc}^3 \times \left\{
\begin{array}{ll}
   10^{-0.4 \,( 1.41  \,a^{1/4})}       & \quad\quad a \le 3.1\E{pc} \\
   10^{-0.4 \,( 29.7  \,a^{1/4}-6.68)}  & \quad\quad 3.1\E{pc} < a \le 20\E{pc} \\
   10^{-0.4 \,( 10.3  \,a^{1/4}+0.61)}  & \quad\quad a > 20\E{pc}                       
\end{array}
\right.
 \label{eq.bulge_rho}
\end{equation}
with $a \equiv 0.57 z^2+\sqrt{0.57^2 z^4 + x^2+y^2+ 1.11 z^2}$ in kpc,

\begin{equation}
  \rho_\M{disk}(x,y,z)  =  0.27 \Msun/\M{pc}^3 \times \exp{\left(-\frac{\sqrt{x^2+y^2}}{h_\sigma}\right)} \,\M{sech}^2{\left(\frac{z}{h_z}\right)} ,
 \label{eq.disk_rho}
\end{equation}
with $h_\sigma\equiv6.4\E{kpc}$ and $h_z\equiv0.3\E{kpc}$,

\begin{equation}
  \rho_\M{halo}(x,y,z) = 0.065 \Msun/\M{pc}^3 \times
  \frac{1}{1+\left(r/r_\M{c}\right)^2}\quad\quad  r\le 100\E{kpc}
\end{equation}

with $r \equiv (x^2+y^2+z^2)^{1/2}$ and $r_\M{c}=4\E{kpc}$.

Integrating over these density profiles gives a total mass 
for the bulge of $4.44\times10^{10}\Msun$, for the disk of 
$4.18\times10^{10}\Msun$ and for the halo of $122.7\times10^{10}\Msun$.

We used a bulge and disk inclination of $i=77^\circ$, a position
angle of the disk major axis of $P.A.=38^\circ$ and
$P.A.=50^\circ$ for the bulge.

For the  transversal velocity distribution we assumed
\begin{equation}
  \pv(\vt) =  \frac{1}{\sigls^2}
  \,\vt 
  \,\exp{\left(- \frac{\vt^2+v_0^2}{2\sigls^2}  \right)}
\, I_{0}{\left(\frac{v_0\,\vt}{\sigls^2}\right)}   ,
\end{equation}
with $\sigls = [\sigl^2+(\sigs \Dol/\Dos)^2]^{0.5}$.
The velocity dispersions are 
    $\sigma_\M{bulge}= 100\E{km\ s}^{-1}$,
    $\sigma_\M{disk} =  30\E{km\ s}^{-1}$,
    $\sigma_\M{halo} = 166\E{km\ s}^{-1}$,
    $\sigma_\M{MW-halo} = 156\E{km\ s}^{-1}$.
The additional rotations are taken into account as
$v_0(\Dos,\Dol,v_\M{rot,l},v_\M{rot,s},v_{\odot-\M{M31}})$, where we used
$v_\M{rot,bulge}=30\E{km\ s}^{-1}$ and 
$v_\M{rot,disk}=235\E{km\ s}^{-1}$ and an observer's
motion of $v_{\odot-\M{M31}}=129\E{km\ s}^{-1} (\Dos-\Dol)/\Dos$.

\section{Event rate}
\label{app.eventrates}

Eq.~\ref{eq.eventrate} contains the following analytical functions [we have
dropped the variables on the right-hand-side (rhs) mostly; all
functions on this side can be expressed as a function of the variables
on the left-hand-side (lhs)]
\begin{displaymath}
  \renewcommand{\arraystretch}{1.0}
  \begin{array}{rcl}
\Upsilon(\DF/\F+1)                           & = & \Upsilon(\A)   = 2\sqrt{ u{\left(\frac{\A+1}{2}\right)^2- {u(\A)}^2}} = \sqrt{8} \,\frac{\left[(\A+1)^{3/2}-\A(\A+3)^{1/2}\right]^{1/2}}{[(\A-1)(\A+1)(\A+3)]^{1/4}} \\
\Psi(\DF/\F+1)                               & = & \dudA \, \Upsilon^2(\A) =  4 \sqrt{2} \,\frac{\left[\A+(\A^2-1)^{1/2}\right]^{1/2} \,\left[(\A+1)^{3/2}-\A(\A+3)^{1/2}\right]}{(\A^2-1)^{7/4} (\A+3)^{1/2}} \\
D_\M{ol}^\ast(\Rstar,\Dos,\ml,\DF,\F,)         & = & \Dos \left(1 + \frac{\DF(2\F+\DF)}{C\,\Dos} \right)^{-1} \quad\M{with}\quad C:=\frac{16 \,\F^2 \,G\ml}{c^2 \,\Rstar^2}\\
u_0^\ast(\Rstar,\Dol,\Dos,\ml)                   & = & \,\left(2\left(1+\left(\frac{\Rstar \,\Dol}{2 \,\RE \,\Dos}\right)^2\right)^{1/2}-2\right)^{1/2} \\ & & \M{with} \quad  \RE :=\frac{\sqrt{4G\ml}}{c} \sqrt{\frac{\Dol (\Dos-\Dol)}{\Dos}}\\
\Omega^\ast(\DF,\Dos,\F,\ml,\Rstar)               & = &  \left|\frac{dD_\M{ol}^\ast(\DF)}{d\DF}\right| = 2 C \Dos^2 (\F + \DF) \left(C \Dos + \DF(2\F+\DF)\right)^{-2} \\ 
\Upsilon^\ast(\uub,\Rstar,\Dol,\Dos,\ml)   & = & 2 \sqrt{ \frac{2 (A_0^\ast+1)}{\sqrt{ (A_0^\ast-1)(A_0^\ast+3)}} - 2 - \uub^2} \quad\M{with}\quad  A_0^\ast = \sqrt{1+\left(\frac{2 \,\RE \,\Dos}{\Rstar \,\Dol}\right)^2}\\
\vt(\Dol,\Dos,\ml,\tfwhm,\DF,\F)             & = & \frac{\RE(\Dol,...)}{\tfwhm} \,\Upsilon \\
v_\M{t}^\ast(\uub,\Dol,\Dos,\ml,\tfwhm,\Rstar) & = & \frac{\RE}{\tfwhm} \,\Upsilon^\ast(\uub,\dots)\\
\end{array}
\end{displaymath}
and the color-magnitude relations of stars, $\pcmd(\Mlum,\Col)$, the
mass function of stars and potential MACHOs, $\xi(\ml)$, the spatial
density of {\bf sources} $n(x,y,\Dos)$ on the line of sight, the mass
density of {\bf lenses}, $\rho(x,y,\Dol)$, and the transversal
projected lens-source velocity distribution $\pv(\vt)$.  The mass
density of lenses depends on the dynamical model and the number density
of sources is constrained by the observed light of the stellar
population. If the extinction value of the stellar population changes,
or equivalently, the $(M/L)$ of the stellar population is
changed, the event rate scales linearly (and not quadratically).
\\
The first term inside the brackets of the rhs of
Eq.~\ref{eq.eventrate} collects contributions from events that do
not show finite source signatures, i.e.  events with impact parameters
larger than the projected source radius, $\uub>\ufs\approx \Rstar\Dol
\ (2\RE\Dos)$ (see Eqs. 65 and 66 in \cite{2006ApJS..163..225R}).  The
second term on the rhs is due to events showing finite source
signatures, i.e. events with impact parameters closer than the
projected source radius, $\uub<\ufs$, we used Eq.~67 in
\cite{2006ApJS..163..225R}.
\\
All functions with {``$\ast$''} depend on the radius
$\Rstar(\Mlum,\Col)$.

\section{Lens mass probability distribution}
\label{sec.analy_massprob}

The event rate differential in Eq.~(\ref{eq.eventrate}) depends on 7
arguments: The first 5 arguments ($x$, $y$, $\tfwhm$, $\DF$, $\Col$)
are observables of lensing events; the brightness of the source stars,
$\Mlum$, is not directly observable in light curves (the
color-luminosity distributions of stars can instead be statistically
described using observations of resolved stellar populations in M31 or
using stellar population models); the sixth argument, the lens mass,
$\ml$, is the quantity to be statistically constrained (lens mass
functions and amplitudes) from observing lensing events.  One can
integrate Eq.~(\ref{eq.eventrate}) to obtain lower order differentials:
\begin{equation}
\renewcommand{\arraystretch}{1.0}
  \begin{array}{lr}
		\hspace{4cm} & \hspace{6.3cm} \\
\frac{d^6 \Gamma(x, y, \tfwhm, \DF, \Col ,  \ml)}{\,dx \,dy \,d\tfwhm
   \,d\DF  \,d\Col  \,d\ml} 
   = 
& 
\int \frac{d^7 \Gamma(x, y, \tfwhm, \DF,  \Col , \ml, \Mlum)}{\,dx \,dy
   \,d\tfwhm \,d\DF \,d\Col \,d\ml \,d\Mlum }  \,d\Mlum 
\\
  \end{array}
\label{eq.eventrate_6dim}
\end{equation}
\begin{equation}
\renewcommand{\arraystretch}{1.0}
  \begin{array}{lr}
		\hspace{4cm} & \hspace{6.3cm} \\
\frac{d^5 \Gamma(x, y, \tfwhm, \DF, \Col )}{\,dx \,dy \,d\tfwhm
   \,d\DF  \,d\Col  } 
   =  
&
\int \int 
\frac{d^7 \Gamma(x, y, \tfwhm, \DF,  \Col , \ml, \Mlum)}{\,dx \,dy
   \,d\tfwhm \,d\DF \,d\Col \,d\ml \,d\Mlum }  \,d\Mlum  \,d\ml
\\
  \end{array}
\label{eq.eventrate_5dim}
\end{equation}
\begin{equation}
\renewcommand{\arraystretch}{1.0}
  \begin{array}{lr}
		\hspace{4cm} & \hspace{6.3cm} \\
\frac{d^4 \Gamma(x, y, \tfwhm, \DF)}{\,dx \,dy \,d\tfwhm
   \,d\DF } 
   =  
&
\int \int \int 
\frac{d^7 \Gamma(x, y, \tfwhm, \DF,  \Col , \ml, \Mlum)}{\,dx \,dy
   \,d\tfwhm \,d\DF \,d\Col \,d\ml \,d\Mlum }  \,d\Mlum   \,d\Col  \,d\ml \\
  \end{array}
\label{eq.eventrate_4dim}
\end{equation}
In the following we assume, that one can measure the location of an
event without any error.  Some of the remaining observables ($\tfwhm,
\DF, \Col$) might have fairly large errors. Let $p(o,o^\M{meas})$ be
the probability for measuring $o^\M{meas}$ with $o$ being the true
value. We then can estimate the error-weighted values of the
differentials in Eqs. \ref{eq.eventrate_6dim}, \ref{eq.eventrate_5dim}
and \ref{eq.eventrate_4dim} at the location of the observables:

\begin{itemize}
\item without color measurement 
\begin{equation}
\renewcommand{\arraystretch}{1.7}
  \begin{array}{lr}
		\hspace{6cm} & \hspace{6cm} \\
\left\langle
{\frac{d^4 \Gamma(x, y, \tfwhm, \DF)}{\,dx \,dy \,d\tfwhm
   \,d\DF } }
\right\rangle
_{x^\M{meas},y^\M{meas}} ( \tfwhm,
   \DF )
   := 
&
\frac{d^4 \Gamma(x^\M{meas}, y^\M{meas}, 
\tfwhm, \DF)}{\,dx \,dy \,d\tfwhm
   \,d\DF } 
  \end{array}
\label{eq.eventrate_obs_4dim_1}
\end{equation}
\begin{equation}
\renewcommand{\arraystretch}{1.7}
  \begin{array}{lr}
		\hspace{6cm} & \hspace{6cm} \\
    \multicolumn{2}{l}{
\left\langle
{\frac{d^4 \Gamma(x, y, \tfwhm, \DF)}{\,dx \,dy \,d\tfwhm
   \,d\DF } }
\right\rangle
_{x^\M{meas},y^\M{meas},\tfmeas,
   \DF^\M{meas} }
   :=  
    } 
    \\
    \multicolumn{2}{r}{
\int \int
\frac{d^4 \Gamma(x^\M{meas}, y^\M{meas}, 
\tfwhm, \DF)}{\,dx \,dy \,d\tfwhm
   \,d\DF } 
\,p(\tfwhm,\tfmeas)
\,p(\DF,\DFmeas)
\,d\tfwhm \,d\DF  
 }
  \end{array}
\label{eq.eventrate_obs_4dim_2}
\end{equation}
\begin{equation}
\renewcommand{\arraystretch}{1.7}
  \begin{array}{lr}
		\hspace{6cm} & \hspace{6cm} \\
    \multicolumn{2}{l}{
    \left\langle
    {\frac{d^5 \Gamma(x, y, \tfwhm, \DF ,  \ml)}{\,dx \,dy \,d\tfwhm
       \,d\DF    \,d\ml} }
    \right\rangle
    _{x^\M{meas},y^\M{meas},\tfmeas,\DFmeas} 
     (\ml)  :=   
    } 
    \\
    \multicolumn{2}{r}{
    \int \int \int 
    \frac{d^5 \Gamma(x^\M{meas}, y^\M{meas}, 
    \tfwhm, \DF,  \ml)}{\,dx \,dy \,d\tfwhm
       \,d\DF \,d\ml} 
   \,p(\tfwhm,\tfmeas)
   \,p(\DF,\DFmeas)
   \,d\tfwhm \,d\DF  
    }
  \end{array}
\label{eq.eventrate_obs_5dim_3}
\end{equation}
\item with color measurement $\Colmeas$
\begin{equation}
\renewcommand{\arraystretch}{1.7}
  \begin{array}{lr}
		\hspace{6cm} & \hspace{6cm} \\
\frac{d^2 \Gamma(\tfwhm, \DF)}{d\tfwhm\,d\DF}:= &\\
\left\langle
{\frac{d^5 \Gamma(x, y, \tfwhm, \DF, \Col)}{\,dx \,dy \,d\tfwhm
   \,d\DF  \,d\Col} }
\right\rangle
_{x^\M{meas},y^\M{meas},\Colmeas}
       (\tfwhm, \DF)
   :=
&
\int 
\frac{d^5 \Gamma(x^\M{meas}, y^\M{meas}, 
\tfwhm, \DF, \Col )}{\,dx \,dy \,d\tfwhm
   \,d\DF  \,d\Col} 
\,p(\Col,\Colmeas) \,d\Col 
  \end{array}
\label{eq.eventrate_obs_5dim_1}
\end{equation}
Eq.~\ref{eq.eventrate_obs_5dim_1} can be transformed
(To transform a distribution $df(x)/dx$ to $dg(u)/du$
  with $u(x)$ and its inverse $x(u)$ we used
$  dg(u)/du=df[x(u)]/dx \left|dx(u)/du\right|$,
see also footnote~\ref{fn.mag_flux}) into
\begin{equation}
  \frac{d^2 \tilde{\Gamma}(\vartheta,\delta_m)}{d\vartheta\,d\delta_m} = 
  0.4\;\ln(10)^2\;10^\vartheta\;\fluxR{\left(\delta_m\right)}\;
  \frac{d^2 \Gamma{\left(10^\vartheta, \fluxR{\left(\delta_m\right)}\right)}}{d\tfwhm\,d\DF}
\label{eq.eventrate_obs_5dim_log}
\end{equation}
 defining
$\vartheta\equiv\log_{10}{\tfwhm}$ and $\delta_m\equiv \magR(\DF)$.
\begin{equation}
\renewcommand{\arraystretch}{1.7}
  \begin{array}{lr}
		\hspace{6cm} & \hspace{6cm} \\
    \multicolumn{2}{l}{
\left\langle
{\frac{d^5 \Gamma(x, y, \tfwhm, \DF, \Col)}{\,dx \,dy \,d\tfwhm
   \,d\DF  \,d\Col} }
\right\rangle
_{x^\M{meas},y^\M{meas},\Colmeas,
     \tfmeas, \DFmeas }
   :=  
    } 
    \\
    \multicolumn{2}{r}{
\int \int \int
\frac{d^5 \Gamma(x^\M{meas}, y^\M{meas}, 
\tfwhm, \DF, \Col )}{\,dx \,dy \,d\tfwhm
   \,d\DF  \,d\Col} 
\,p(\tfwhm,\tfmeas)
\,p(\DF,\DFmeas)
\,p(\Col,\Colmeas)
\,d\tfwhm \,d\DF \,d\Col 
    }
  \end{array}
\label{eq.eventrate_obs_5dim_2}
\end{equation}
\begin{equation}
\renewcommand{\arraystretch}{1.7}
  \begin{array}{lr}
		\hspace{6cm} & \hspace{6cm} \\
    \multicolumn{2}{l}{
    \left\langle
    {\frac{d^6 \Gamma(x, y, \tfwhm, \DF, \Col ,  \ml)}{\,dx \,dy \,d\tfwhm
       \,d\DF  \,d\Col  \,d\ml} }
    \right\rangle
    _{x^\M{meas},y^\M{meas},\tfmeas,\DFmeas, \Colmeas} 
     (\ml)  :=   
    } 
    \\
    \multicolumn{2}{r}{
    \int \int \int 
    \frac{d^6 \Gamma(x^\M{meas}, y^\M{meas}, 
    \tfwhm, \DF, \Col ,  \ml)}{\,dx \,dy \,d\tfwhm
       \,d\DF  \,d\Col  \,d\ml} 
   \,p(\tfwhm,\tfmeas)
   \,p(\DF,\DFmeas)
   \,p(\Col,\Colmeas)
   \,d\tfwhm \,d\DF \,d\Col 
    }
  \end{array}
\label{eq.eventrate_obs_6dim}
\end{equation}
\end{itemize}
Equation~(\ref{eq.eventrate_obs_6dim}) gives the contribution to the
event rate (with event characteristics as observed) as a function of
lens mass.  The probability for a lens with mass $\ml$ causing an
observed event can therefore be written as
\begin{equation}
  \hat{p}(\ml) \propto  
\left\langle
{\frac{d^6 \Gamma(x, y, \tfwhm, \DF, \Col ,  \ml)}{\,dx \,dy \,d\tfwhm
   \,d\DF  \,d\Col  \,d\ml} }
\right\rangle
_{x^\M{meas},y^\M{meas},\tfmeas,\DFmeas, \Colmeas}(\ml)
\quad . 
\label{eq.mass_probability}
\end{equation}
We will evaluate the lens mass probability function for each
self lensing (bulge-bulge, bulge-disk, disk-bulge and disk-disk) and
halo (M31-halo-bulge, M31-halo-disk, MW-halo-bulge, MW-halo-disk) lensing
scenario. 

Equations~\ref{eq.eventrate_obs_5dim_1} and
\ref{eq.eventrate_obs_4dim_1} describe the distributions of events (at
the location of the observed event) in the flux-excess-lensing
time-scale plane, for the case where the observed color of the event
is used (Eq.~\ref{eq.eventrate_obs_5dim_1}) and for all events,
irrespective of their color (Eq.~\ref{eq.eventrate_obs_4dim_1}).
Studying the event distribution in this plane is very useful, since
one can immediately see, if an event is unlikely (given a theoretical
model), and how data quality (imposing limits on detectable flux
excess and time scale) restricts the measurable event rate.

The relative values of Eqs.~\ref{eq.eventrate_obs_5dim_2} and
\ref{eq.eventrate_obs_4dim_2} give the relative probabilities for the
different lensing scenarios, if the color of the event is used or not
used, respectively.

\acknowledgments{We are very grateful to the anonymous referee for a
  lot of constructive suggestions. This research was supported by the
  Son\-der\-for\-schungs\-be\-reich SFB 375 of the Deut\-sche
  For\-schungs\-ge\-mein\-schaft (DFG) and by the DFG cluster of
  excellence Origin and Structure of the Universe
  (www.universe-cluster.de).}

\end{document}